\documentstyle[preprint,aps,graphicx]{revtex}
\input{psfig.sty}

\begin{document}
\draft 

\title{Solid-phase structures of the Dzugutov pair potential}

\author{J. Roth, Institut f\"ur Theoretische und Angewandte Physik,
  Universit\"at Stuttgart, Paffenwaldring 57, 70550 Stuttgart, Germany\\
A. R. Denton, Department of Physics, Acadia University, Wolfville, NS, 
  Canada B0P 1X0}

\date{\today}

\maketitle

\begin{abstract}
In recent computer simulations of a simple monatomic system
interacting via the Dzugutov pair potential, freezing of the fluid 
into an equilibrium dodecagonal quasicrystal has been reported 
[M. Dzugutov, Phys. Rev. Lett. {\bf 70}, 2924 (1993)].
Here, using a combination of molecular dynamics simulation and 
thermodynamic perturbation theory, we conduct a detailed analysis
of the relative stabilities of solid-phase structures of the 
Dzugutov-potential system.  
At low pressures, the most stable structure is found to be a bcc 
crystal, which gives way at higher pressures to an fcc crystal.  
Although a dodecagonal quasicrystal and a $\sigma$-phase crystal 
compete with the bcc crystal for stability, they remain always metastable.
\end{abstract}

\bigskip



\section{Introduction}

Several years ago~\cite{Dzugutov1}, Dzugutov introduced a new model 
pair potential for the purpose of studying glass transitions by means of 
molecular dynamics (MD) simulation.  Most simulations of glass transitions 
to date have been performed using binary mixtures, since one-component 
simple liquids, when supercooled, readily nucleate crystallites.
The Dzugutov potential, which features a pronounced maximum at a range
typical of next-nearest-neighbor coordination distances in close-packed 
crystals, suppresses crystallization by construction, and thus facilitates
glass formation.  Following its initial successful use in studies of 
supercooled liquids~\cite{Dzugutov1,Dzugutov2}, the Dzugutov potential 
was subsequently adopted in simulations of freezing~\cite{Dzugutov3,Dzugutov4}.  
Contrary to expectations, however, the observed solid structure 
was determined to be, remarkably, a monatomic dodecagonal quasicrystal.
The structure, also known as tetrahedrally or topologically close packed 
(tcp)~\cite{Daams}, is of the Frank-Kasper type and is composed of layers 
of square-triangle tilings.  In previous work, the Dzugutov potential 
and the associated dodecagonal structure have been used to study
self-diffusion in quasicrystals~\cite{self-diffusion}.

The motivation for the present work stems primarily from our interest in 
the nucleation and stability of quasicrystal phases.  Further motivation 
comes from the realm of colloid physics, where, given the extreme tunability 
of colloidal interparticle interactions, it is conceivable that a 
Dzugutov-like pair potential might be engineered to produce bulk samples 
of one-component quasicrystals~\cite{Denton-Lowen}.  The main purpose
of the study reported here is to chart, by means of both MD simulation 
and thermodynamic perturbation theory, the fluid-solid phase diagram 
of the Dzugutov-potential system.  

The paper is organized as follows.  In Section \ref{simul}, after specifying
the pair potential, we give details of the simulation methods, 
and describe our analyses of the resulting solid structures.  
Section \ref{theory} outlines the theoretical methods used, while 
Section \ref{struct} characterizes the quasicrystal and
other structures of interest.  In Section \ref{cogro} we present results -- 
from both simulation and theory -- for relative stabilities 
of competing solid structures.  
Among our main results, we find that
(1) at low temperatures and pressures, the stable solid structure is the 
bcc crystal; (2) at high pressures, the stable solid is the fcc crystal; and
(3) the stable ground state ($T=0$) is never the dodecagonal quasicrystal, 
but rather a periodic crystal whose structure depends on the pressure.
Finally, in Section \ref{discus} we summarize and discuss implications 
of the results for future work. 

\section{Molecular Dynamics Simulations}\label{simul}

\subsection{The Interaction}

The Dzugutov pair potential\cite{Dzugutov2}, plotted in Fig. 1, is defined by

\begin{equation}
\Phi(r)~=~\Phi_1(r)~+~\Phi_2(r), 
\label{defpot-a}
\end{equation}
where
\begin{equation}
\Phi_1(r)~=~\left\{ \begin{array}
{l@{\quad\quad}l}
A(r^{-m}-B)\exp\left(\frac{\displaystyle c}{\displaystyle r-a}\right), & r<a \\
0, & r\geq a, \\
\end{array} \right.
\label{defpot-b}
\end{equation}
and
\begin{equation}
\Phi_2(r)~=~\left\{ \begin{array}
{l@{\quad\quad}l}
B\exp\left(\frac{\displaystyle d}{\displaystyle r-b}\right), & r<b \\
0, & r\geq b, \\
\end{array} \right.
\label{defpot-c}
\end{equation}
with the parameters:
\begin{center}
\begin{tabular}{c|c|c|c|c|c|c}
m & A & c & a & B & d & b\\
\hline
16 & 5.82 & 1.1 & 1.87 & 1.28 & 0.27 & 1.94\\
\end{tabular}
\end{center}

The potential is characterized by a
minimum at $r=1.13$ $\sigma$ of depth $-0.581$ $\epsilon$, having 
the same form as that of the Lennard-Jones potential, followed by 
a maximum at $r=1.63$ $\sigma$ of height $0.460$ $\epsilon$. 
The maximum is designed to prevent the system from crystallizing into 
simple crystal structures. Beyond the maximum the potential tends to 
zero continuously and is cut off at a range of $r_{\rm c}=1.94$ $\sigma$, 
which ensures that CPU times remain within reasonable limits.

\subsection{Simulation Method}

Classical isothermal (NVT) and isothermal-isobaric (NPT) MD simulations 
were performed using the constraint method\cite{allen87}. 
An extension of this method allows us to introduce 
constant temperature or pressure gradients\cite{lancon}. 
Newton's equations of motion were integrated using a fourth-order
Gear predictor-corrector algorithm (see, {\it e.g.}, \cite{allen87}) 
with a time increment of $\delta t=0.0005$ $\sigma\sqrt{m/\epsilon}$ 
for all simulations. 
Periodic boundary conditions were applied to an orthorhombic
simulation cell, whose volume (in the NPT simulations) was permitted
to change isotropically.

The sample sizes range from 54 to 1024 atoms, with the lengths of the cell 
along the three orthogonal coordinate axes being chosen to make 
the sample shapes as close to cubic as possible. 
Simulations with the stable phases were carried out with samples
containing 54, 250, and 1024 atoms for bcc; 60 and 480 atoms for the
$\sigma$-phase; and 108 and 500 atoms for fcc. 
Most results are reported for 250, 500, and 1024 atoms, though if not 
explicitly stated, the sample size is irrelevant.
The potential energy and enthalpy per atom were found to be 
quite insensitive to sample size.

\subsection{Structural Analysis}\label{analys}

To analyze the structures that arise in a simulation at finite temperature, 
we quenched the system by setting the temperature in our NPT-MD program 
to zero, thereby using the program as a steepest descent algorithm. 
Quenching forces the system to seek out its local energy minimum, 
which facilitates structure identification.

Dzugutov\cite{Dzugutov2} has pointed out that vacancies may play a
major role in stabilizing quasicrystalline structures.  To determine 
the number of vacancies in a sample, we begin by constructing 
the Voronoi cells and their dual, the Delaunay cells, and determining 
from the latter the distribution of free volumes in the structure.
Although the free volume distribution gives a reasonable representation 
of the interstitial sites, it greatly overestimates the vacancies 
(by a factor of about ten).  This is because the Delaunay cells are 
face-to-face packed tetrahedra, whereas the vacancies should be represented 
by spheres that could easily cover several tetrahedra. 
To determine the correct number of vacancies, we first select the
Delaunay cells that are large enough to accommodate an atom and do not 
overlap with an already existing atom. 
With these cells we create trees of mutually  
overlapping cells, the nodes of the tree being the centers of the
Delaunay cells and the edges the distance vectors
between cells too close to be filled simultaneously. 
We then fill the tree with spheres, starting at the outermost ends. 
After adding a sphere, all the Delaunay cells connected to its node are 
discarded. The next sphere is then added onto the next outermost node 
remaining. The procedure is repeated until the whole tree is filled, 
after which the algorithm repeats with another Delauney cell not belonging 
to the current tree.
This method allows us to fill the sample as densely as possible with 
vacancies.

In order to help characterize and distinguish the solid structures 
observed in the simulations, we have computed from the atomic coordinates 
both radial and angular distribution functions. 
However, such averaged functions often do not allow unique identification 
of a structure, which requires as well the spatial distribution of bonds. 
Thus, we have also generated bond order diagrams.  These are 
stereographic projections of the nearest-neighbor bonds constructed 
as follows.  First, all neighbor vectors are determined
and normalized to unit length.  Next, the vectors are placed at a common
origin so that their endpoints lie on the unit sphere.  Finally, the
distribution of the points on the sphere is represented by
stereographic projections along the three coordinate axes. 
The pictures thus obtained reveal the global symmetry of the sample.

\section{Theory}\label{theory}

For comparison with the simulation data, we have independently calculated
the phase behavior of the system by means of thermodynamic perturbation theory. 
Taking the Dzugutov pair potential as input, we apply the approximate
theory of Weeks, Chandler, and Andersen (WCA)~\cite{WCA} to a classical 
system of $N$ pair-wise-interacting particles in a volume $V$.
The WCA approach is especially well suited to pair potentials that
contain a steeply repulsive core and has been successfully applied 
to the Lennard-Jones potential~\cite{CA1,Mederos,Rascon}, which has 
a repulsive core similar to that of the Dzugutov potential.

The WCA approximation splits the pair potential $\phi(r)$ 
at its first minimum into a short-range repulsive reference potential 
$\phi_0(r)$ and a perturbation potential $\phi_{\rm p}(r)$ and 
prescribes a mapping of the reference system onto an 
effective hard-sphere (HS) system.
The Helmholtz free energy $F$ of the system separates correspondingly 
into reference and perturbation parts, according to
\begin{equation}
F=F_0+\int_0^1{\rm d}\lambda \langle\Phi_{\rm p}\rangle_{\lambda},
\label{pert1}
\end{equation}
where $F_0$ is the free energy of the reference system, 
\begin{equation}
\Phi_{\rm p}\equiv\sum_{i<j}\phi_{\rm p}(|{\bf r}_i-{\bf r}_j|)
\label{Phip}
\end{equation}
is the total perturbation energy, and $\langle\cdots\rangle_{\lambda}$
denotes averaging with respect to the probability distribution of a system 
with pair potential $\phi_{\lambda}(r)=\phi_0(r)+\lambda\phi_{\rm p}(r)$.
Expansion of $\langle\Phi_{\rm p}\rangle_{\lambda}$ in powers of $\lambda$ 
about the reference system ($\lambda=0$) generates an exact 
perturbation series.
Mapping the reference system onto an effective HS system, the free energy 
may be expressed, to {\it first-order} in the perturbation potential, as
\begin{equation}
F[\rho({\bf r})]~=~F_{\rm HS}[\rho({\bf r})]
~+~\frac{2\pi N^2}{V}\int_0^{\infty}{\rm d}r'~r'^2~
g_{\rm HS}(r';[\rho({\bf r})])~\phi_{\rm p}(r'),
\label{pert2}
\end{equation}
where $F_{\rm HS}[\rho({\bf r})]$ and $g_{\rm HS}(r;[\rho({\bf r})])$ are 
the free energy and radial distribution function (rdf), respectively, 
of the HS reference system, both functionals of the 
equilibrium one-particle number density $\rho({\bf r})$.  
The rdf is defined, in turn, according to
\begin{equation}
g_{\rm HS}(r;[\rho({\bf r})])\equiv \frac{1}{4\pi\rho^2 V}\int{\rm d}\Omega
\int{\rm d}{\bf r}'\rho^{(2)}({\bf r}',{\bf r}'+{\bf r}),
\label{rdf1}
\end{equation}
as an orientational and translational average of the 
two-particle density $\rho^{(2)}({\bf r},{\bf r}')$.
The second- and higher-order terms are proportional to successively 
higher powers of inverse temperature $1/T$, the coefficients being
related to mean fluctuations of $\Phi_{\rm p}$~\cite{HM}.
Accuracy of the first-order approximation [Eq.~(\ref{pert2})] is thus
assured as long as fluctuations in $\Phi_{\rm p}$ remain sufficiently small 
relative to the thermal energy $k_{\rm B}T$.

The free energy of the fluid phase is calculated via the 
uniform limit ($\rho({\bf r}) \longrightarrow \rho$) of Eq.~(\ref{pert2}), 
using the essentially exact Carnahan-Starling and Verlet-Weis 
forms~\cite{HM} for the HS free energy per particle, $f_{\rm HS}(\rho)$,  
and rdf, $g_{\rm HS}(r)$, respectively.
For the solid phase, the HS free energy functional is approximated 
by means of classical density-functional (DF) theory~\cite{DF}.
The DF approach is based on the existence of a functional 
${\cal F}[\rho({\bf r})]$ of the density $\rho({\bf r})$ that 
satisfies a variational principle, according to which 
${\cal F}[\rho({\bf r})]$ is minimized -- for given 
average density and external potential -- by the equilibrium density, 
its minimum value equaling the Helmholtz free energy $F$.
In the absence of an external potential, ${\cal F}[\rho({\bf r})]$ may be
decomposed into an exactly known ideal-gas contribution 
\begin{equation}
{\cal F}_{\rm id}[\rho({\bf r})] = k_{\rm B}T\int{\rm d}{\bf r}\rho({\bf r})
\Bigl[\ln(\rho({\bf r}) \Lambda^3)-1\Bigr],
\label{Fid1}
\end{equation}
which is the free energy in the absence of interactions
($\Lambda$ being the thermal de Broglie wavelength) and an excess 
contribution ${\cal F}_{\rm ex}[\rho({\bf r})]$, depending entirely upon 
internal interactions. 

Here we approximate the excess free energy of the HS solid by the 
modified weighted-density approximation (MWDA)~\cite{MWDA,CA2}, 
which gives a reasonable description of the HS system.
The MWDA maps the excess free energy per particle of the solid onto that 
of a corresponding uniform fluid of effective density, according to
\begin{equation}
\frac{1}{N}{\cal F}_{\rm ex}^{MWDA}[\rho({\bf r})] = f_{\rm HS}(\hat\rho),
\label{MWDA}
\end{equation}
where the effective (or weighted) density
\begin{equation}
\hat\rho \equiv \frac{1}{N} \int d{\bf r} \int d{\bf r}'
\rho({\bf r})\rho({\bf r}')w(|{\bf r}-{\bf r}'|;\hat\rho)
\label{rhohat}
\end{equation}
is a self-consistently determined weighted average of $\rho({\bf r})$.  
The weight function $w(r)$ is specified by normalization 
and by the requirement that ${\cal F}_{\rm ex}^{MWDA}[\rho({\bf r})]$ generate 
the exact two-particle (Ornstein-Zernike) direct correlation function 
$c(r)$ in the uniform limit.  
This leads to an analytic relation~\cite{MWDA} between $w(r)$ and 
the fluid functions $f_{\rm HS}$ and $c(r)$, computed here using the 
solution of the Percus-Yevick (PY) integral equation 
for hard spheres~\cite{HM}.  

Practical calculation of $F_{\rm HS}[\rho({\bf r})]$ and 
$g_{\rm HS}(r;[\rho({\bf r})])$ requires specifying the solid density, 
{\it i.e.}, the coordinates of the lattice sites (equilibrium particle 
positions) and the shape of the density distribution about these sites. 
Here we consider fcc, hcp, bcc, and $\sigma$-phase crystals and the 
dodecagonal quasicrystal structures described below in Sec.~\ref{struct}.
The density distribution is modelled by the Gaussian ansatz.  This
places at each site ${\bf R}$ a normalized isotropic Gaussian, such that
\begin{equation}
\rho({\bf r})=\left(\frac{\alpha}{\pi}\right)^{3/2}\sum_{\bf R}
\exp(-\alpha|{\bf r}-{\bf R}|^2),
\label{Gauss}
\end{equation}
the single parameter $\alpha$ determining the width of the distribution. 
The Gaussian ansatz has been shown by simulation~\cite{YA_OLW} to 
reasonably describe the density distribution of close-packed crystals.
For nonoverlapping neighboring Gaussians -- consistently the case here 
-- the ideal-gas free energy per particle [Eq.~(\ref{Fid1})] is 
very accurately approximated by
\begin{equation}
\frac{1}{N}{\cal F}_{\rm id} = \frac{3}{2}k_{\rm B}T\ln(\alpha\Lambda^2)-
\frac{5}{2},
\label{Fid2}
\end{equation}
to within an irrelevant constant.
The HS free energy is obtained, for a given solid structure and average
density, by minimizing the approximate functional 
${\cal F}_{\rm HS}[\rho({\bf r})]={\cal F}_{\rm id}[\rho({\bf r})]+
{\cal F}_{\rm ex}^{MWDA}[\rho({\bf r})]$ [from Eqs.~(\ref{MWDA}), 
(\ref{rhohat}), and (\ref{Fid2})] with respect to $\alpha$.
Predictions of the MWDA for free energies and pressures of HS solids are 
in good agreement with simulation data for both fcc~\cite{MWDA} and 
bcc~\cite{CR} crystals.
Although the theory underpredicts, by roughly 20 \%, the Lindemann ratios
(ratio of root-mean-square particle displacement to nearest-neighbor distance)
at melting for both crystal symmetries, it is only the free energies that
determine thermodynamic phase behavior.  Note that in simulations of 
the HS bcc crystal a constraint of single-cell occupancy is usually 
imposed in order to stabilize the crystal against shear.

The perturbation free energy in Eq.~(\ref{pert2}) requires knowledge 
of the hard-sphere rdf, which may be expressed, in general, as a sum 
over coordination shells:
\begin{equation}
g_{\rm HS}(r;[\rho({\bf r})])=\sum_{i=1}^{\infty}g^{(i)}(r;[\rho({\bf r})]).
\label{rdf2}
\end{equation}
The functionals $g^{(i)}(r;[\rho({\bf r})])$ are obtained using the 
approach of Rasc\'on {\it et al}.~\cite{Rascon}, which approximates
the second and higher coordination shells in mean-field fashion and 
corrects the first coordination shell for nearest-neighbor correlations.  
Thus, ignoring correlations for $i\geq 2$, and substituting 
$\rho^{(2)}({\bf r},{\bf r}')=\rho({\bf r})\rho({\bf r}')$,  
together with Eq.~(\ref{Gauss}), into Eq.~(\ref{rdf1}), yields
\begin{equation}
g^{(i)}(r;[\rho])=\frac{1}{4\pi\rho}(\frac{\alpha}{2\pi})^{1/2}~
\frac{n_i}{r R_i}\exp[-\alpha(r-R_i)^2/2],\qquad i\geq 2,
\label{rdf3}
\end{equation}
where $n_i$ is the coordination number and $R_i$ the lattice vector 
magnitude of the $i$th shell.  The first peak is parametrized by
\begin{equation}
g^{(1)}(r;[\rho({\bf r})])=\frac{A\exp[-\alpha_1(r-r_1)^2/2]}{r},
\qquad r\geq d,
\label{rdf4}
\end{equation}
where $d$ is the effective HS diameter and where the parameters $A$, 
$\alpha_1$, and $r_1$ are determined by imposing three sum rules, 
namely the virial equation (relating the contact value to the bulk 
pressure $P$), normalization to the nearest-neighbor coordination number 
$n_1$, and approximation of the first moment by its mean-field value.
Together then, the HS pressure $P=\rho^2\partial f_{\rm HS}/\partial\rho$
and the value of $\alpha$ that minimizes $F_{\rm HS}[\rho({\bf r})]$ 
determine $g_{\rm HS}(r;[\rho({\bf r})])$ and so the 
perturbation free energy for a given solid structure.
The approximation expressed by Eqs.~(\ref{rdf2})-(\ref{rdf4}) is in 
excellent agreement with simulation data for the HS fcc crystal, 
and has been successfully applied, in a perturbation theory, to 
Lennard-Jones and square-well solids~\cite{Rascon}.
The approximation also has been tested against, and found to closely match, 
Monte Carlo simulation data for $g_{\rm HS}(r;[\rho({\bf r})])$ of a HS bcc 
crystal (at density $\rho\sigma^3=1.1$), subjected to a single-cell occupancy 
constraint to suppress shear instability~\cite{Watzlawek}.
Further simulations will be required, however, to test the approximation 
for the HS bcc crystal at lower densities, where next-nearest-neighbor
correlations and anisotropies in the density distribution 
may not be negligible.

It remains still to specify the effective HS diameter $d$.
According to the WCA prescription, $d$ is the root of the 
nonlinear equation
\begin{equation}
\int{\rm d}{\bf r}y_{\rm HS}(r;[\rho({\bf r})];d)\Delta e(r) = 0,
\label{d}
\end{equation}
where $y_{\rm HS}(r;[\rho({\bf r})];d)\equiv\exp[\phi_{\rm HS}(r;d)/k_{\rm B}T]
g_{\rm HS}(r;[\rho({\bf r})];d)$ is the HS cavity function and 
\begin{equation}
\Delta e(r) = \exp[-\phi_0(r)/k_{\rm B}T]-\exp[-\phi_{\rm HS}(r;d)/k_{\rm B}T]
\label{e}
\end{equation}
is a function that is nonzero only over a narrow range $\xi d$ 
($\xi \ll 1$) around $r=d$. 
This choice ensures that the free energy of the reference system 
differs from that of the effective HS system only by terms of 
$O(\xi^4)$ and higher.
In practice, lacking knowledge of the cavity function of the HS solid
for $r<d$, we expand the quantity $r^2 y_{\rm HS}(r;[\rho({\bf r})];d)$
in a Taylor series about $r=d$ and retain the first three terms.

The theory set out above provides a reasonable approximation for 
the Helmholtz free energy of the system at temperatures of order 
$\epsilon/k_{\rm B}$ and higher.  
From the free energy, any bulk thermodynamic property then
may be calculated.  Of particular relevance to phase behavior are the 
pressure and chemical potential.  In Sec.~\ref{theopred}, 
we present our theoretical predictions for the phase diagram of the 
Dzugutov-potential system.

\section{Solid Structures}\label{struct}

\subsection{Dodecagonal Quasicrystal}
\label{deco}\label{basflip} 

The structural model of the dodecagonal quasicrystal that we have
investigated is a layered system that is periodic in one direction,
but quasiperiodic and twelve-fold symmetric in the perpendicular
plane \cite{roth96}. It is of Frank-Kasper type, {\it i.e.}, it is mostly 
tetrahedrally close-packed, and can be described as a periodic
$ABA\bar B$ stacking of a primary dodecagonal layer $A$ and two
secondary hexagonal layers, $B$ and $\bar B$, which are rotated by
$30^\circ$ with respect to each other to obtain dodecagonal
symmetry. The atoms in layer $A$ form the vertices of a simple tiling
made of squares, triangles, $30^\circ$ rhombi and two kinds of
hexagons. The threefold symmetric hexagon is known as the
``shield''. These tiles, together with their decorations, are shown in 
Fig.~\ref{basictile}.   A sample of a square-triangle configuration is
displayed in Fig.~\ref{tiling}. The dodecagonal quasicrystal structure
also can be regarded as the decoration of a simple dodecagonal tiling
\cite{gaehler88,beeli90}.

The stability of the monatomic Frank-Kasper-type decoration of the
square-triangle-rhombi-shield tiling with the potential of Eq.\
(\ref{defpot-a}) was reported by Dzugutov \cite{Dzugutov3}. 
Upon cooling below the glass transition temperature, a glass
forms, which transforms, after a very long annealing time, into a
dodecagonal quasicrystal. The underlying tiling structure is mainly a
decorated square-triangle tiling with a few rhombi and shields.

The aperiodicity of quasicrystals forbids periodic boundary conditions 
in the simulation. Taking a finite patch
with open boundary conditions also should be avoided, as the
surface energy would affect structural stability. 
A solution is to use periodic approximants, which are finite,
orthorhombic cells whose boundaries match on opposite sides.
In this way, periodic boundary conditions may be applied, 
as is done throughout this paper.

\subsection{Square-Triangle Crystals}\label{cryst}

In addition to the quasicrystalline tilings, it is also possible to
generate crystalline phases with squares and triangles decorated
in the same fashion as for the quasicrystals. If only squares are used,
the A15- or $\beta$-W-phase (also known as cP8 Cr$_{3}$Si) is obtained,
whereas a pure triangle tiling results in the Z structure (hp7
Zr$_3$Al$_4$). If both squares and triangles are permitted, the
$\sigma$-phase or $\beta$-U (also known as tp30 Cr$_{46}$Fe$_{54}$)
and the H-phase are obtained. The two phases differ in the arrangement
of the tiles. The vertex configurations of the crystalline phases are 
shown in Fig.~\ref{crystvert}. They are denoted by a, z, h, $\sigma$
according to the phase in which they appear. 

The unit cell of the $\sigma$-phase can be subdivided into two regular
triangles and two squares. The atoms at the vertices of these tiles
are 14-fold coordinated, while the atom in the center of the triangles 
is 15-fold coordinated. 
The remaining atoms on the edges and in the interior of the square are
12-fold coordinated icosahedra.  Since all coordination shells have
a triangular surface, all atoms are tetrahedrally close packed.
Pure square-triangle structures and tilings with additional shields
are very stable\cite{roth96}.  Rhombi, however, are unstable and 
transform into the other tiles.


\section{Results}\label{cogro}

\subsection{Analysis and Comparison of Solid Structures}\label{geomet}

In this section we discuss the geometric properties of the amorphous 
structure, the bcc and fcc crystals, the nucleated tcp-phase, and 
the $\sigma$-phase.
We have used three diagnostics to compare the structures: the
radial distribution function (rdf), the angular distribution function (adf)
of nearest neighbors separated by a distance less than the first minimum 
of the rdf (usually $r<1.6$ $\sigma$), and bond order diagrams, 
which aid in identifying the global symmetry even if the symmetry elements 
are oriented in random directions. 

\subsubsection{Radial Distribution Functions}

Figure~\ref{rdftran}a compares the radial distribution functions of 
the structures most commonly observed in the simulations.  
Typical of the amorphous structures is 
an asymmetric first peak, which appears to consist of two overlapping shells, 
followed by a second maximum in the range $1.7<r<2.7$ $\sigma$, which is 
the well-known double peak.  The unusually sharp slope on the short-distance 
side indicates a well-defined second-nearest-neighbor distance, which is 
caused by the repulsive part of the maximum of the potential. 
The maximum at about
1.9 $\sigma$ is formed by two opposite corners of a bipyramid consisting of two
regular tetrahedra. A second remarkable sharp slope is found at the
third maximum at about 2.7 $\sigma$.  Notice that the rdfs of the tcp 
structures obtained on cooling are completely indistinguishable from 
the amorphous case. 

The rdf of the $\sigma$-phase shows an example of the perfect 
square-triangle-rhombi-shield phases.
The quasicrystalline and other crystalline and approximant phases
differ only in the fine structure of the subpeaks.  Evidently
the tcp-phase rdf is a broadened envelope of the $\sigma$-phase rdf.

The rdfs of the bcc and fcc structures, however, are radically different. 
For bcc, the first maximum is indeed split, and in the second maximum 
the short-distance part is {\em lower} than the next peak. 
This maximum is now formed by the
distances across the tetragonal octahedron in the bcc structure.
At higher temperatures, when the peaks of the rdf are broadened and
overlap, the rdf of the bcc-phase is similar to the rdf of the
tcp-phases except that the weights of the two subpeaks of the second
maximum between 1.7 $\sigma$ and 2.7 $\sigma$ are interchanged. 

The rdf of the fcc crystal exhibits a peak at about 1.6 $\sigma$. 
Since this is the position of the potential maximum, 
it is clear that the fcc-phase is unstable at low pressures. 
Figure \ref{rdftran}b shows the transition from a bcc-phase to
the fcc-phase at $k_{\rm B}T/\epsilon=0.75$. 
The small maximum at 1.5 $\sigma$ indicates formation of regular squares, 
which are characteristic of fcc. 
The MD data are strikingly similar to theoretical predictions 
for the hard-sphere fcc crystal (Fig.~\ref{rdftran}c).

The sequence of phases at low temperatures and increasing pressure
becomes clear when we examine the rdfs: for bcc and the $\sigma$-phase, 
the first two overlapping atomic shells occupy the minimum of the
potential. The next shell is beyond the maximum. For fcc, the first
shell is also in the minimum and the second shell is at the maximum. If
the structures are compressed, the energies of bcc and the $\sigma$-phase
{\em increase}, since the second maximum of the rdf moves up the maximum
of the potential. The energy of fcc {\em decreases}, since the second 
maximum of the rdf moves down the potential maximum.  

\subsubsection{Angular Distribution Functions}

The results for the angular distribution functions (Fig.~\ref{winplot}) 
are consistent with the results for the rdfs.  
The amorphous and tcp structures are indistinguishable. 
The tcp-phase adf is a broadened version of the adf
of the $\sigma$-phase. All of these phases, as well as the liquid show
two maxima: a rather narrow extremum at small angles around
60$^{\circ}$ and a broad peak at about 120$^{\circ}$. Both maxima
indicate the existence of equilateral triangles. The adfs of
bcc and fcc are again completely different. Especially remarkable
are the maxima at an angle of 90$^{\circ}$, indicating the existence
of rectangles in these phases. For bcc these angles are
formed by distances between atoms along the four-fold axis.

\subsubsection{Bond Order Diagrams}

We have seen that the amorphous and tcp structures obtained
by cooling cannot be distinguished by the rdf and adf. 
Are they really different? 
As shown in Fig.~\ref{bodbcc}, bond order diagrams can give a clear answer.
The diagram for bcc (Fig.~\ref{bodbcc}a) is relatively simple, exhibiting
seven maxima -- four from the separation along the space diagonals and three
from the distances along the four-fold directions.  As the sample
shown here was not perfect, thin bridges join the maxima. 
The liquid (not shown) is characterized by an isotropic distribution of the
bonds covering the whole sphere homogeneously.  The diagram for the
tcp-phase (Fig.~\ref{bodbcc}b) is somewhat more complicated, featuring
an equator with twelve maxima indicating the presence of a quasicrystal. 
Other prominent features are two further circles of maxima at
higher latitudes and two peaks at the poles. For non-perfect samples
the distribution of the maxima is distorted, and the symmetry may not
be dodecagonal.  Occasional ring-like arrangements of overlapping maxima
surrounded by further maxima suggest twinning and multi-grain samples. 

Comparing samples classified as amorphous or
tcp-phases, we find that a continuous transition between the two may be
possible. In Fig.~\ref{bodbcc}b the maxima can be seen quite 
clearly.  In other samples, however, the maxima are almost obscured by
a rather homogeneous background noise. Quenching and
annealing improves the diagrams only marginally.  It is possible
that some of the amorphous samples actually consist of a number of
micro-grains.  If in fact the case, this would mean that 
the amorphous and tcp-phases have the same {\em local} arrangement of atoms,  
although the amorphous sample did not succeed in ordering globally.

\subsubsection{Real-Space Representation of the Structures}

Real space pictures (snapshots) of the samples also help to distinguish 
the structures.
In most cases, the bcc samples look quite defect-free, with only the
vacancies visible. Sometimes we find two differently oriented
domains in the simulation box.  Liquid samples obviously do not
show any regularities. In the quenched amorphous structures, however,
there are sometimes partially ordered parts, underlining our claim
that they contain micrograins.
The tcp samples, as exemplified by Fig.~\ref{bodbcc}c, have layered 
structures, which, if viewed perpendicular to the layers, resemble
the perfect sample in Fig.~\ref{tiling}, characterized by
centered ring-like structures formed by the 14- and 15-fold
coordinated atoms.  The quality of the pictures is often low, however, 
as the samples may be twinned or contain several grains.

\subsection{Ground-State Structures}\label{ground}\label{resim}

The equilibrium structure of a specific potential at a given
temperature and pressure may be determined, in principle, by a global
minimization of the Gibbs free energy $G$. In practice, however, this
procedure is not feasible with MD simulations, since direct transitions 
between local minima are rarely observed.
Even if special methods are used to switch between closely related
structures like bcc and fcc, there still may exist a free energy
barrier high enough to prevent a transition\cite{kremer}.
Instead of attempting to minimize $G$, one identifies promising structures,
computes the thermodynamic functions, and compares them.
An alternative is thermodynamic integration of $P$, starting at $P=0$
and integrating towards higher pressures.

At $T=0$, where entropy no longer affects stability, identifying
the ground state simplifies considerably.
Here the Helmholtz free energy $F$ equals the internal energy $U$,  
which in turn equals the potential energy $E_{\rm pot}$, since the 
kinetic energy vanishes.
Furthermore, the Gibbs free energy $G$ becomes equal to the enthalpy $H$ 
and the pressure $P$ is determined by the virial equation, since the kinetic
pressure $k_{\rm B}T$ also vanishes.  Common tangent constructions on
curves of $U(V,T=0)$ vs. $V$ yield the stability ranges of competing phases 
and curves of enthalpy $H(P,T=0)$ vs. $P$ intersect at 
the phase coexistence pressures.

We have calculated the ground-state energies by two independent methods.
First, taking the perfect structures, we have computed lattice summations
of the Dzugutov pair potential. 
Second, starting from a perfect structure, we have relaxed the system 
with the MD simulation program in an isothermal-isobaric ensemble, setting 
$k_{\rm B}T/\epsilon=0.001$ and $P$ to the desired value. In this mode, 
molecular dynamics acts as a steepest-gradient optimization algorithm. 
The pressure is derived from the virial equation. 
In contrast to a lattice sum calculation, where only the volume is scaled, 
in the simulations all atoms may move independently.  
Therefore the results may (and do) differ slightly from 
the lattice sum calculations for perfect structures. 

Since the Dzugutov potential [Eq.\ (\ref{defpot-a})] is isotropic and has
a single minimum, it should favor densely packed structures if the volume 
is not restricted. 
An optimal packing in three dimensions would consist of 
regular tetrahedra, but such a packing does not exist. Now there are
two choices to solve this dilemma: either introduce other coordination
polyhedra, as in fcc crystals, or use irregular tetrahedra, as in tcp phases. 
In the Frank-Kasper phases the coordination polyhedra are additionally
restricted to deltahedra with five or six triangles
meeting at a vertex. This condition is fulfilled for the icosahedron,
and certain polyhedra with 14, 15 and 16 vertices. Although bcc is
tetrahedrally close-packed, it is not a Frank-Kasper phase since its
coordination polyhedron (a rhombic dodecahedron) has vertices where
only four triangles meet. 

In a first step towards identifying stable structures we study
stacking variants, distort the phases mentioned above, and examine
various tcp structures.
The fcc structure can be modified by stacking the densely packed layers
differently. We find that the hexagonal close-packed (hcp) and other
stacking variants are considerably less stable than fcc at high
pressures where fcc is more stable than bcc and the $\sigma$-phase.
This result is remarkable since for the Lennard-Jones potential
hcp is known to be slightly more stable than fcc \cite{vdwlj}. 

Distortions of the bcc phase along the principal symmetry
axis always reduce the stability. The same happens for the
$\sigma$-phase if the layer distance is changed from the optimum at
$c/a=1.03$ ($a$ is the edge length of the tiles and c the period along
the z-axis).

The Frank-Kasper phases are of two types: structures
with 16-fold coordinated sites and structures without. The dodecagonal 
quasicrystal, its approximants, and crystalline variants are of the
latter type, called the square-triangle class. Structures containing
16-fold (or higher) coordinated atoms have a lower stability.
The sites with the high coordination numbers are too
numerous and the potential energy increases because of strained bonds. 

We observe the same trend in the square-triangle class. The stability
is lowest for the purely triangular Z-phase since the number of 15-fold sites
is also considerable. The stability increases if the triangles are
separated by squares, but is again rather low if the structure contains
only squares as in the A-phase without 15-fold sites, perhaps because
the A-phase has full cubic symmetry and is therefore more
rigid than the other structures. The $\sigma$-phase, on the other hand, 
is more stable than the H-phase, since it contains only pairs of
triangles instead of rows. More complicated crystalline phases, 
approximants, and the quasicrystals all contain mixtures of squares and
triangles in different arrangements. These structures are all inferior
to the $\sigma$-phase since they must contain larger conglomerates of
triangles. 

In a second step towards identifying ground-state structures, we survey
the published crystallographic structures. From the lists in Refs.\
\cite{westbrook,pearson,deboer}, a variety of structures have been
selected according to the following criteria:  
\begin{enumerate}
\item Coordination numbers between 10 and 15.
\item Derivatives of tcp structures.
\item Derivatives of bcc, especially vacancy-ordered structures. 
\item Quasicrystal approximants.
\item Icosahedral coordination shells.
\end{enumerate}
For each structure examined the required crystallographic data were
taken from Ref.\ \cite{Daams}. A full list of the structures is
given in the Appendix. 

Assembling the results, the following picture emerges at $T=0$
(Fig.~\ref{gs-energy}): the bcc-phase has the absolute minimum potential 
energy at a density of $\rho\sigma^3=0.866$. The $\sigma$-phase acts as the
lower bound for all the square-triangle phases, being minimal at
$\rho\sigma^3=0.879$. The fcc-phase has a potential energy minimum at
$\rho\sigma^3=1.013$.  A common tangent construction shows
that bcc is stable up to $\rho\sigma^3=0.887$, and fcc is stable 
above $\rho\sigma^3=1.057$. 
The relaxed $\sigma$-phase is stable, as determined by MD, only 
within the narrow interval $0.887<\rho\sigma^3<1.057$, whereas the ideal
$\sigma$-phase is never stable, according to a simple lattice sum calculation.
Intersections of the enthalpy curves yield the stability ranges in terms of
pressure.  The most stable structures are bcc for $P\sigma^3/\epsilon<1.70$, 
$\sigma$ for $1.70<P\sigma^3/\epsilon<2.85$, and fcc for 
$P\sigma^3/\epsilon>2.85$.
The sequence of ground-state structures with increasing pressure (and
density) is therefore: bcc -- $\sigma$ -- fcc. The properties of the
various structures are summarized in Table \ref{enemin}.

Vacancy-ordered phases are more stable than pure bcc at densities down
to $\rho\sigma^3=0.7$ (Fig.~\ref{gs-energy}b). In the range 
$0.6<\rho\sigma^3<0.7$ the lowest potential energy is attained by 
a disordered phase formed upon annealing the NiTi$_2$ approximant phase.

\subsection{Finite-Temperature Phases}

From the ground-state calculations we have determined the stable structures 
at $T=0$.  Insight into the topology of the phase diagram at finite
temperatures now can be obtained by observing phase transitions between 
the melt and the solid upon heating, cooling, and compression.
As noted in Sec.~\ref{ground}, it is not possible to determine the 
relative thermodynamic stabilities of two competing phases directly
from conventional MD simulations because of the difficulty of computing
the entropic contribution to the free energy.  However, by observing
the temperature and pressure at which a phase becomes unstable, it is 
possible to establish limits of mechanical stability.  The results of 
our MD stability analysis are consolidated in Fig.~\ref{phasNT}a. 

\subsubsection{Heating Simulations}\label{meltp}

If the bcc and $\sigma$-phase solids are heated at low pressure, the energy 
and enthalpy for bcc remain always lower than those for the $\sigma$-phase. 
At higher pressures, the energy of the $\sigma$-phase drops below that of bcc. 
The differences between the enthalpies at higher temperatures, however, are 
smaller than their fluctuations, such that the relative stabilities of
bcc and the $\sigma$-phase cannot be resolved.

The determination of the melting line has been discussed in
detail in Ref.\ \cite{ptdzu}. Here we present only a brief
summary. The phase transition line was determined by
preparing a solid at $k_{\rm B}T/\epsilon=0$ or 0.4 at fixed pressure and
heating it continuously at rates of $k_{\rm B}\delta T/\epsilon$=0.001 or 
0.002 per timestep until melting was observed. The criterion for melting was 
the divergence of the mean-square displacement. At the same temperature a
sudden rise in the potential energy and an associated drop in the density 
were observed.  Similar simulations have been carried out at constant volume 
starting at $k_{\rm B}T/\epsilon=0.001$ and $P\sigma^3/\epsilon=0.001$.
We emphasize that the transition line thus obtained is not strictly
the equilibrium melting line, since with periodic boundary conditions 
the sample has no surface at which melting could start and
a two-phase coexistence is not possible because the samples are too small.
The solid-fluid transition lines differ only slightly for bcc and the 
$\sigma$-phase. The fcc crystal melts at somewhat higher temperatures. 
At high pressures, however, the
$\sigma$-phase becomes unstable at considerably lower temperatures
than bcc and fcc.

The transition to the fluid also can be determined by expanding a
solid at constant temperature starting from high densities. The 
transition line obtained in this way for bcc crystals is the same as
that determined by heating within the statistical fluctuations
(Fig.~\ref{phasNT}a).
As noted in Ref.\ \cite{ptdzu}, only one fluid phase is observed and 
no transition between a liquid and a vapor phase could be found.

\subsubsection{Cooling Simulations}\label{fspc}

Cooling simulations were carried out in a manner similar to the heating runs.
Starting samples were obtained from solids equilibrated at high temperatures.
The cooling rate was $ k_{\rm B}\delta T/\epsilon=0.002$ per
time step. Similar to melting, the freezing transition is delayed, now
because critical nuclei first must be formed, and subsequent large-scale 
reordering of atoms may be necessary.

We find that the phase nucleating at pressures above $P\sigma^3/\epsilon=5$ 
always has bcc symmetry. If the temperature is lowered to about 
$k_{\rm B}T/\epsilon=0.7-1.0$,
bcc becomes unstable relative to fcc at pressures above
$P\sigma^3/\epsilon=20$. Although a complete transition to fcc cannot be
achieved with our simulation method, we observe a clear indication that
fcc is the preferred structure. In 
the radial distribution function (Fig.~\ref{rdftran}a) we observe the
emergence and growth of a new peak between the first and the second peaks
at about $\sqrt{2}$ times the nearest-neighbor distance, which signals
the formation of regular squares in the close-packed crystal structures. 

Below $P\sigma^3/\epsilon=5$ we do not observe a typical freezing transition 
with a jump in potential energy, but only a sharp kink, reminiscent of 
a glass transition. The nucleating structures are partially ordered and
possess features typical of the tcp structure, namely layering, ring-like
structures, and the Frank-Kasper polyhedra (see Fig.~\ref{bodbcc}c). 
In the following, we refer to such structures as the tcp-phase. 
Although sometimes dodecagonal, the tcp structures often do not have 
a perfect symmetry, and thus may have varying degrees of crystallinity. 

Because of the maximum in the potential, it was not possible in general
to obtain perfect samples. If the pressure is too low,
there is insufficient cohesion to compactify the samples. This is
clearly seen by comparing runs with different cooling rates.  However, 
if equilibrated for a longer time, the samples eventually become much denser.

The density ranges for stability have been obtained by cooling
at constant volume. For the 500-atom sample we obtain the boundary between
the formation of the bcc and the tcp-phase at $\rho\sigma^3=0.87$, 
independent of the cooling rate up to $k_{\rm B}\delta T/\epsilon= 0.0005$ 
per time step. For the 1024-atom sample, however, the boundary is shifted to
$\rho\sigma^3=0.84$ and is observed at the first time for 
$k_{\rm B}\delta T/\epsilon=0.00025$ per time step. 
This is remarkable, since the minimum of the
$\sigma$-phase lies at about $\rho\sigma^3=0.9$.

The formation of the crystalline structures also depends on the sample
sizes and the cooling rates. A sample with 250 atoms and a constant
density of $\rho\sigma^3=0.865$ froze to bcc 
at a cooling rate of $k_{\rm B}\delta T/\epsilon=0.001$. For 500 atoms 
we had to reduce the cooling rate by a factor of one half, and for 1024 atoms
a cooling rate of $k_{\rm B}\delta T/\epsilon=0.00025$ per time step was
necessary to 
obtain a perfect bcc phase, although partial bcc ordering was already
observed at twice this rate.

It is easier to obtain the tcp-phase in a constant
volume simulation (as Dzugutov did) rather than in a constant
pressure ensemble. To some extent, the nucleated structures can be
annealed also at constant volume. However, most of the defects, especially
different domains, cannot be so removed. Annealing at constant pressure
also turns out to be ineffectual.

The transition from fluid to solid also may be observed by compressing
the fluid at a constant pressure gradient of 
$\delta P\sigma^3/\epsilon=\pm0.1$. The transition curve is the same as for 
cooling (Fig.~\ref{phasNT}a), the collapsed structures being again bcc, 
at least for $k_{\rm B}T/\epsilon=0.6$, 0.8, 1.0, 1.5, and 2.5. 

Between the melting and freezing curves we observe a broad hysteresis region, 
within which the thermodynamic phase transition should occur. 
The reason for the broad hysteresis region is the peculiar form of 
the potential.  The maximum strongly inhibits freezing and collapsing 
of the structure, indeed as intended by Dzugutov \cite{lenjon}.

The structures generated by cooling the samples contain free volumes
even if the density or pressure during nucleation is high. In the
case of constant volume cooling the reason is obvious, since the
volume of the nucleating regions shrinks with temperature.  However, 
constant pressure cooling also generates free volumes, 
even at high pressures, since at
the onset of nucleation the frozen domains have a higher density
than the liquid. The rigidity of the solid prevents the simulation box
from contracting fast enough. With the methods described in
Sec.~\ref{analys}, we can show that the free volumes in the ordered
phases are mostly vacancies. 

If the simulation samples are quenched to $T=0$ and $P=0$, 
and the vacancies are filled with atoms, we find that the density of 
bcc rises to $\rho\sigma^3=0.864\pm0.005$, whereas the densities of the
tcp-phase remain at about $\rho\sigma^3=0.847\pm0.005$. 
Although the densities of the bcc samples are close to the ideal value 
of 0.8638, the densities of the tcp-phase are far lower than the ideal value
at the potential energy minimum ($\rho\sigma^3=0.881$).

\subsubsection{Compression Simulations}

If the structures are compressed at fixed temperature, bcc destabilizes first.
One might therefore expect bcc to be stable only at relatively low 
temperatures.  However, this would contradict the cooling simulations 
(Sec.~\ref{fspc}), which yield a bcc structure.  A full picture can be 
obtained only by calculating the Gibbs free energy, since it may be 
kinetically favorable for the system to nucleate bcc crystallites. 

At high pressures the stable structure is clearly fcc, which has the lowest
energy and enthalpy.  Upon compression, this close-packed structure remains 
stable, and radial distribution functions of the decaying bcc and 
$\sigma$-phase structures show new peaks characteristic of fcc.

\subsection{Theoretical Predictions}\label{theopred}

\begin{figure}
\vspace{1cm}
\centerline{\includegraphics[width=8cm]{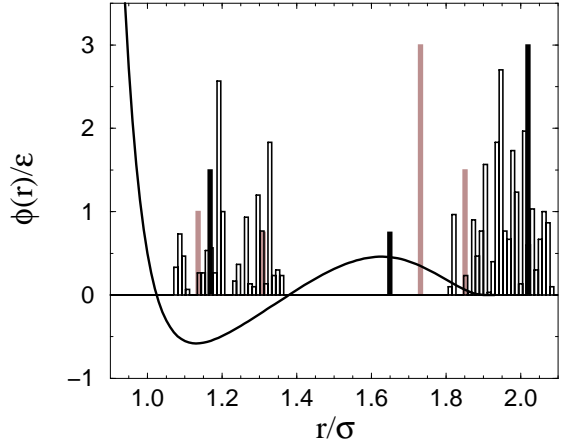}}
\vspace{0.5cm}
\caption[]{
  Dzugutov pair potential together with histograms of neighbor distances for
  fcc (black bars), bcc (gray bars), and $\sigma$-phase (unshaded bars) 
  crystals at reduced density $\rho\sigma^3=0.9$.  Heights of bars 
  are proportional to numbers of neighbors.}
\label{pot}
\end{figure}

\begin{figure}
  \centerline{\includegraphics[width=10cm]{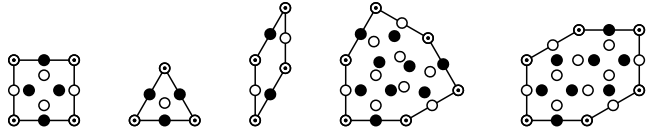}}
\vspace{0.5cm}
\caption[]{The basic tiles of the dodecagonal model: square,
  triangle, rhombus, shield, twofold symmetric hexagon. The dotted
  atoms are placed in $A$-layers $z=1/4$ and 3/4, the white atoms in
  $B$-layers at $z=0$ and the black atoms at $Z=1/2$. All tiles can
  also occur with black and white atoms exchanged, depending on their
  orientation. The twofold symmetric hexagon is unstable and does not
  occur in our tilings.}
\label{basictile}
\end{figure}

\begin{figure}
  \centerline{\includegraphics[width=10cm]{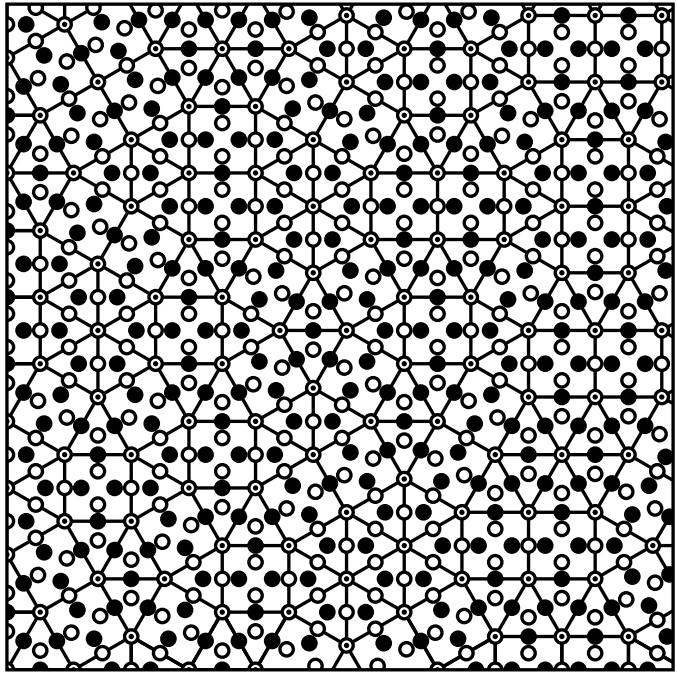}}
\vspace{0.5cm}
\caption[]{Patch of a decorated square-triangle quasicrystal. The
  dotted atoms are at $z=1/4$ and 3/4 in the $A$ layer, the white atoms
  are at $z=0$ in the $B$ layer and the black atoms at $z=1/2$ in the
  $\bar{B}$ layer.}
\label{tiling}
\end{figure}

\begin{figure}
  \centerline{\includegraphics[width=12cm]{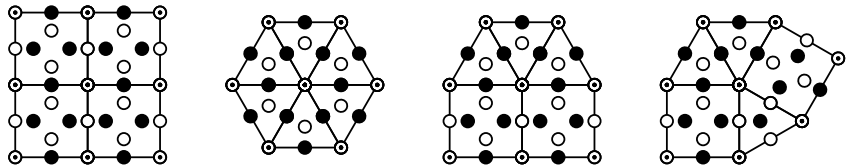}}
\vspace{0.5cm}
\caption[]{Vertex configurations of the crystalline phases with a
  single vertex configuration. From left to right: A-, Z-, H- and
  $\sigma$-phase or a, z, h, and $\sigma$ vertex, respectively.} 
\label{crystvert}
\end{figure}

\begin{figure}
  \centerline{\includegraphics[width=8cm,angle=270]{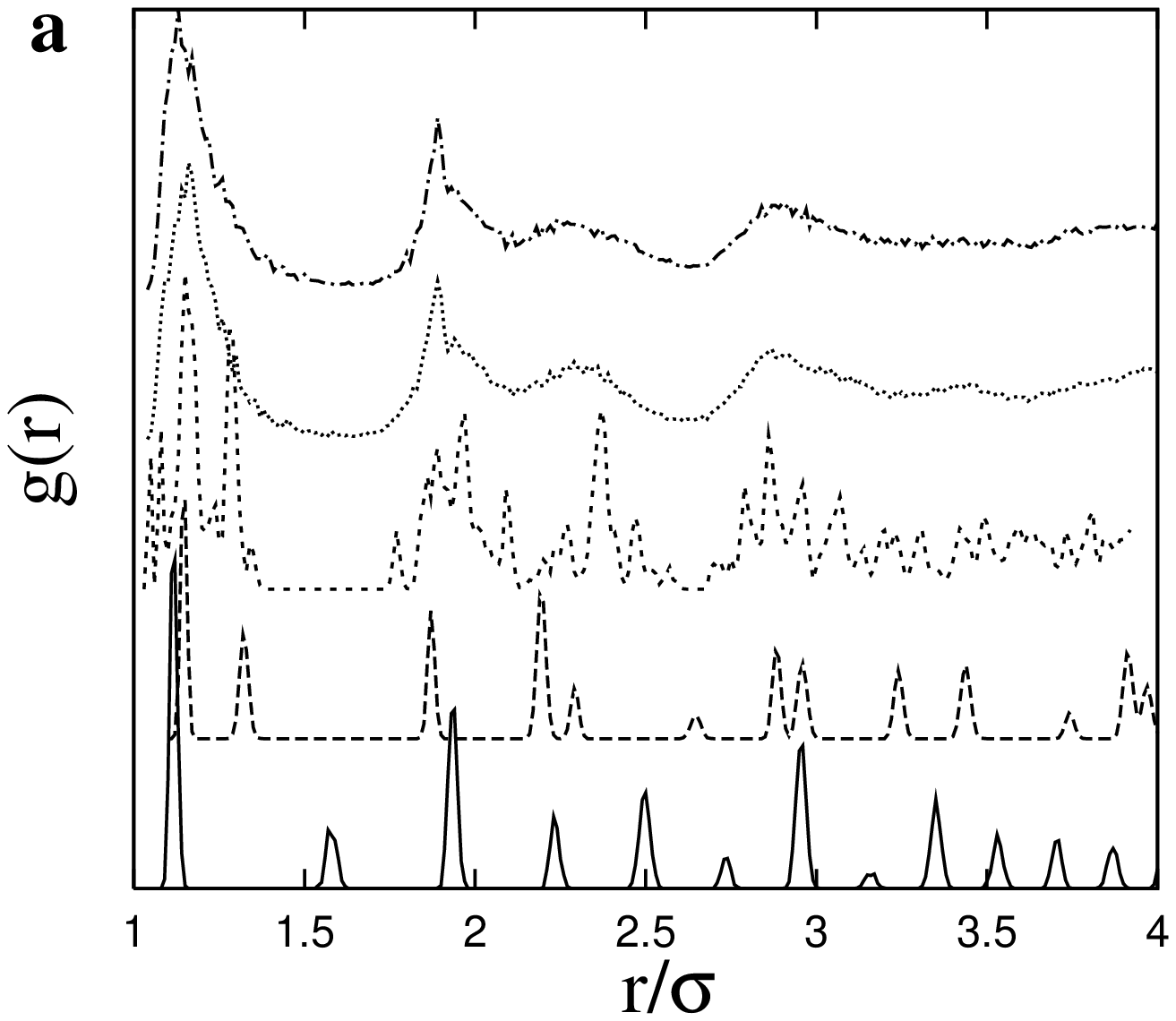}}
\vspace{1cm}
  \centerline{\includegraphics[width=8cm,angle=270]{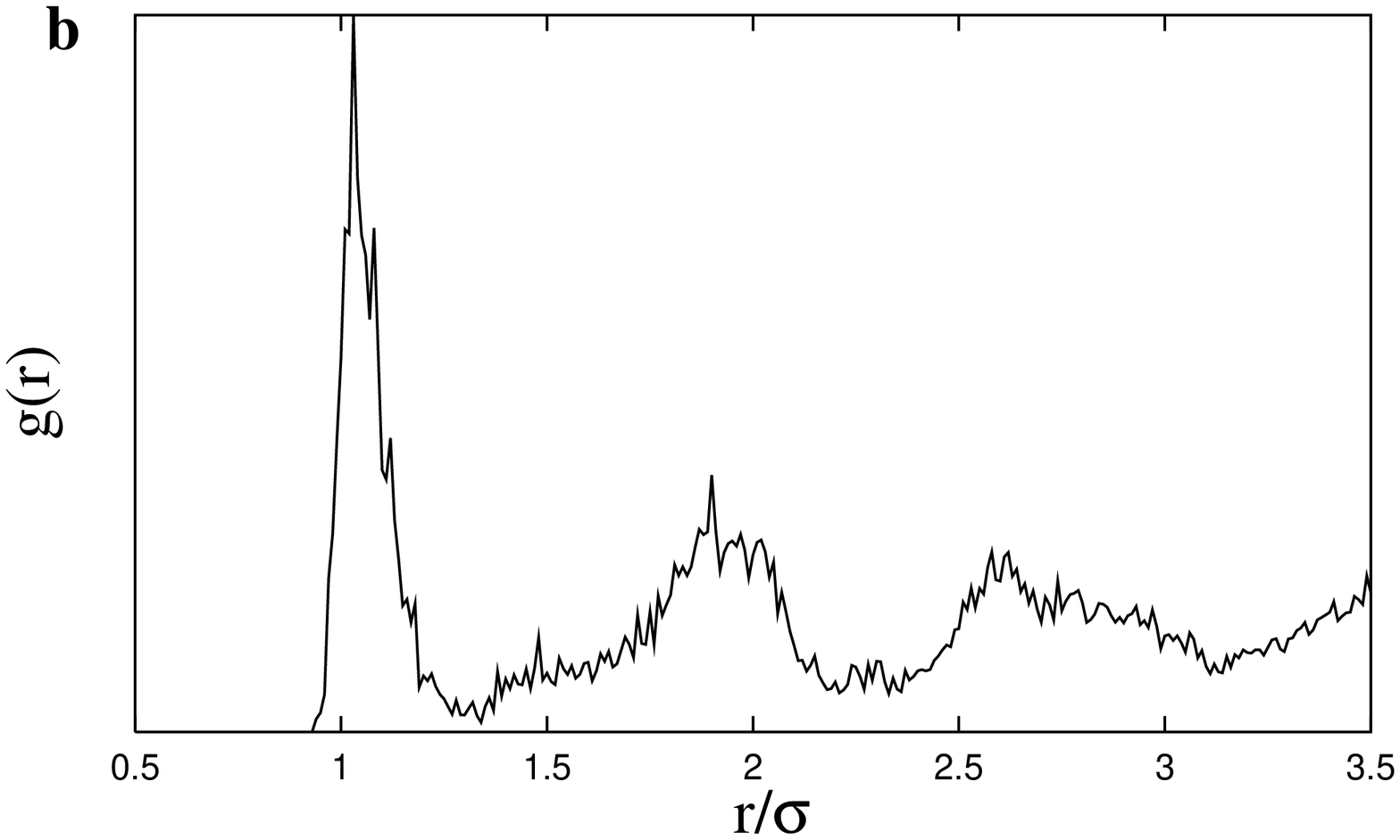}}
\vspace{1cm}
  \centerline{\includegraphics[width=8cm]{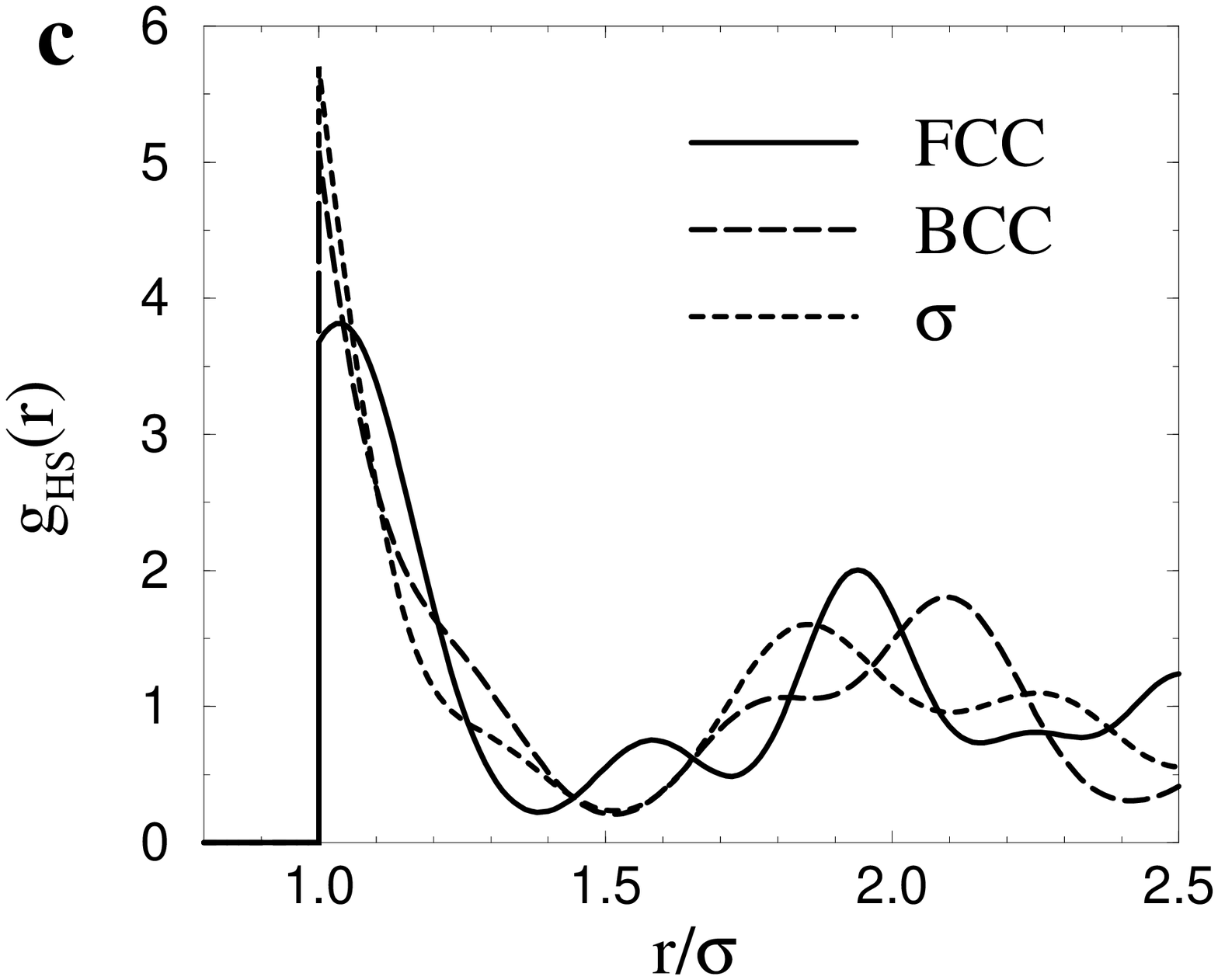}}
\vspace{0.5cm}
\caption[]{Radial distribution functions for various systems.
  (a) From bottom to top: fcc, bcc, $\sigma$-phase, tcp-phase, amorphous phase.
  The samples have been expanded to $P\sigma^3/\epsilon=0.001$ and
  quenched to $T=0$. Vertical scale is in arbitrary units, all curves
  being scaled to the same maximum.
  (b) A sample obtained from cooling at $P\sigma^3/\epsilon=25$ and 
  $k_{\rm B}T/\epsilon=0.75$. The shoulder at $r=1.5$ $\sigma$ indicates 
  the onset of a transition to fcc;
  (c) Hard-sphere solid at reduced density $\rho\sigma^3=1.0$, computed from 
  Eqs. (\ref{rdf2})-(\ref{rdf4}): fcc crystal (solid curve), bcc crystal 
  (long-dashed curve), and $\sigma$-phase (short-dashed curve).}
\label{rdftran}
\end{figure}

\begin{figure}
  \centerline{\includegraphics[width=8cm,angle=270]{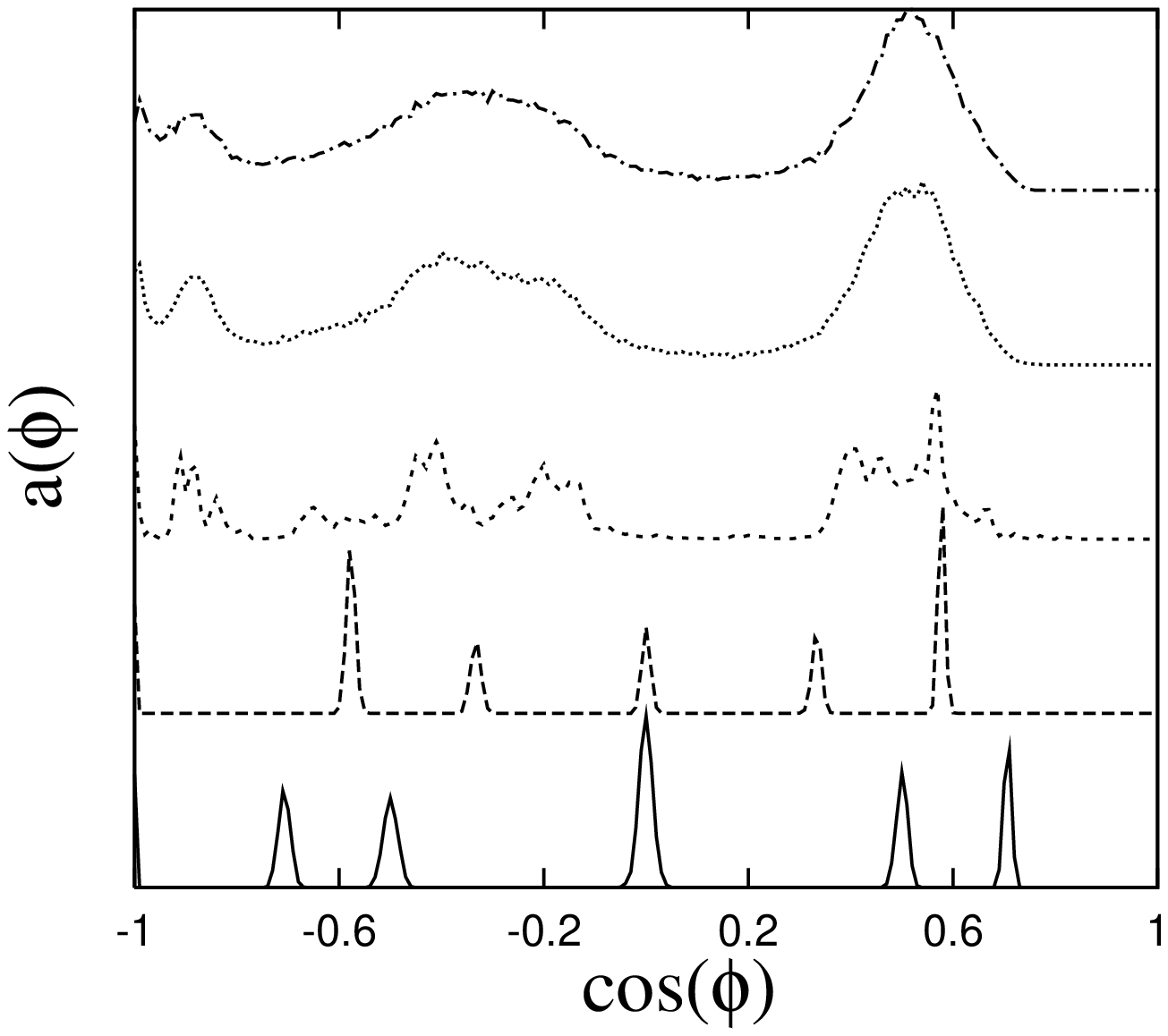}}
\vspace{0.5cm}
\caption[]{Angular distribution functions. From bottom to top: fcc,
  bcc, $\sigma$-phase, tcp-phase, amorphous phase. The samples have
  been expanded to $P\sigma^3/\epsilon=0.001$ and quenched to $T=0$.}
\label{winplot}
\end{figure}

\begin{figure}
  \centerline{\includegraphics[width=8cm,angle=270]{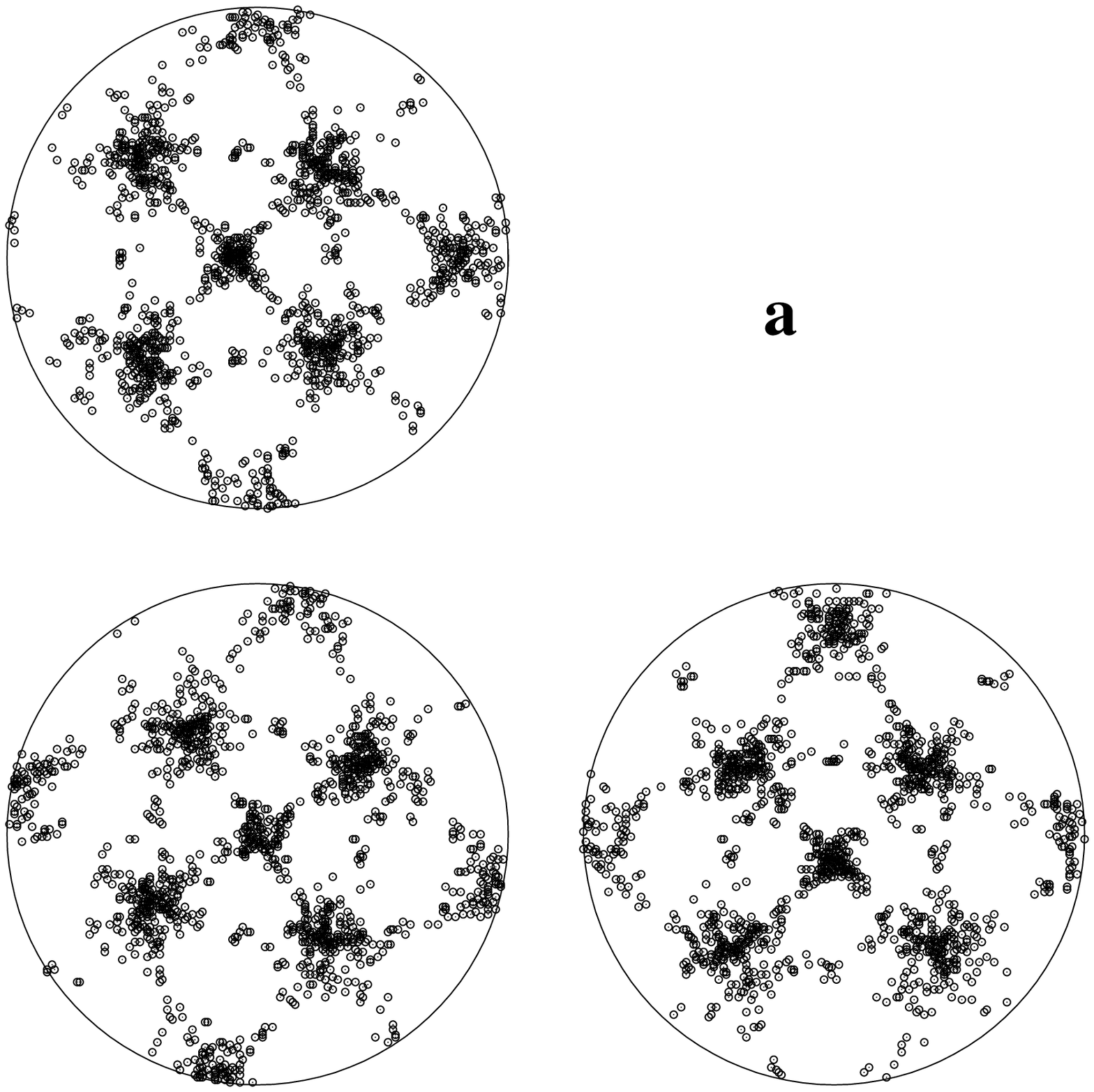}}
\vspace{2cm}
  \centerline{\includegraphics[width=8cm,angle=270]{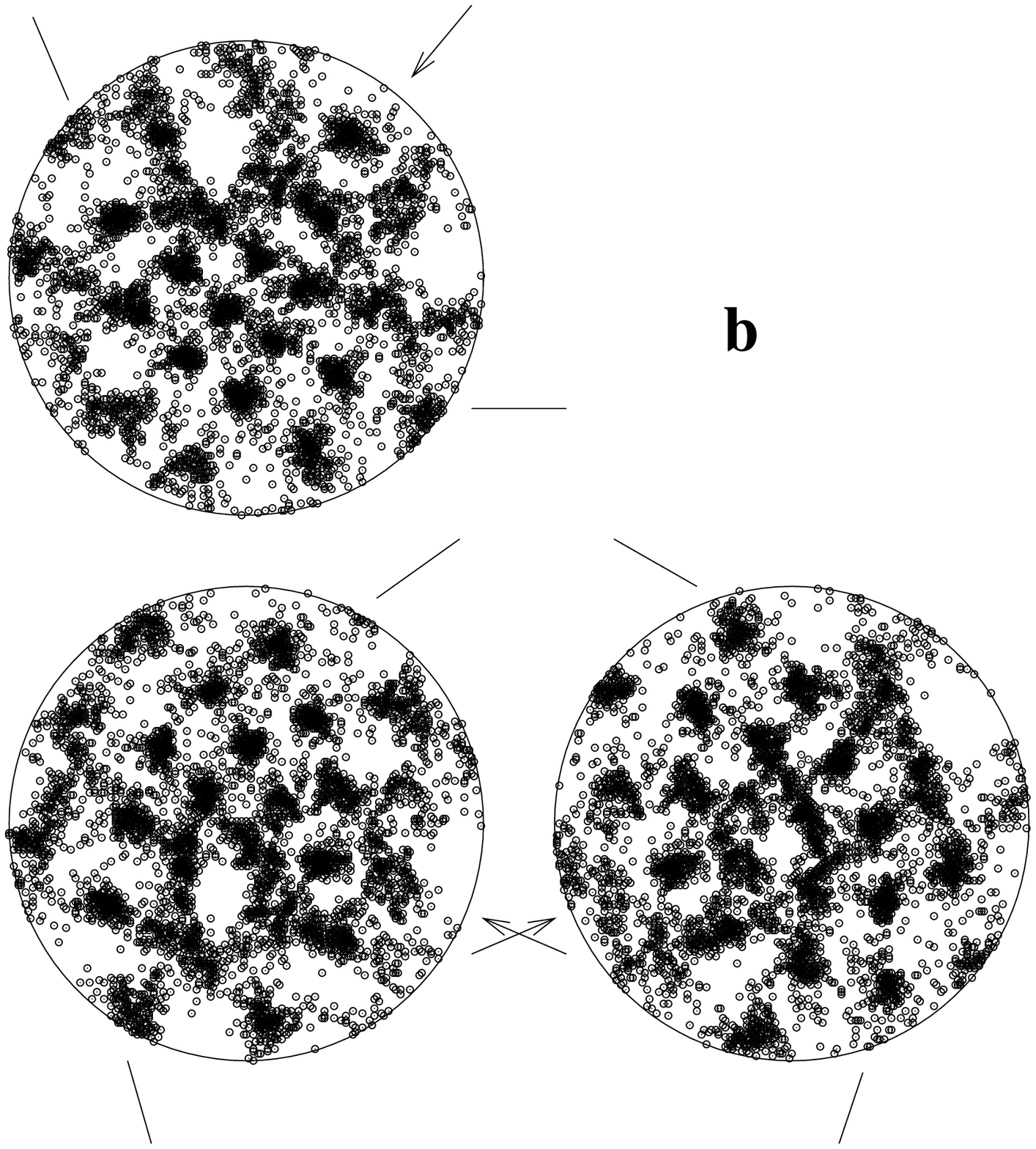}}
\vspace{2cm}
  \centerline{\includegraphics[width=8cm,angle=270]{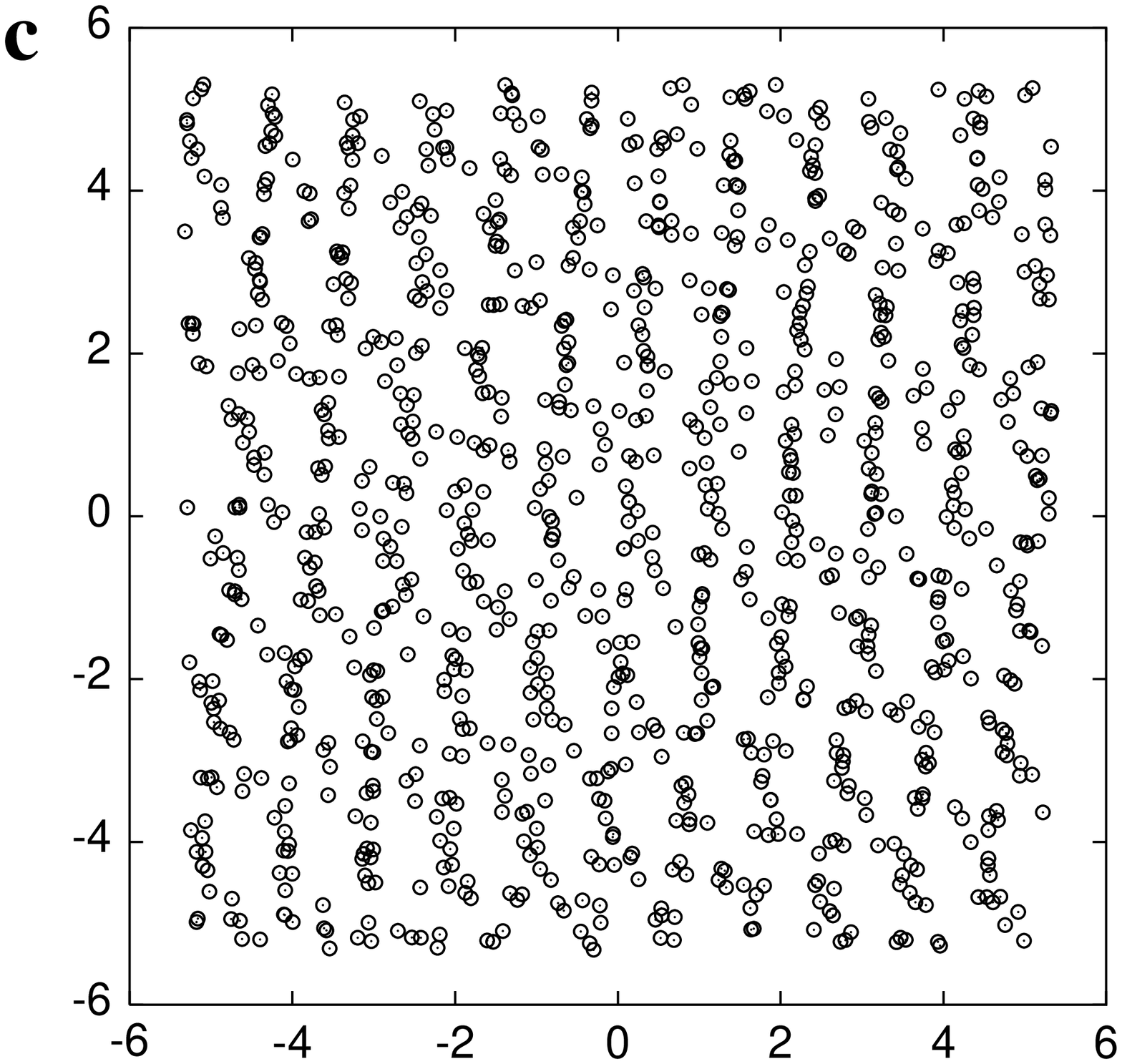}}
\vspace{0.5cm}
\caption[]{
  (a) Spherical projection of the nearest-neighbor vectors of a
  non-perfect bcc sample generated by cooling. Three  views of the
  sphere along perpendicular directions are given;
  (b) Spherical projections of the nearest-neighbor vectors of
  a tcp sample obtained by cooling. Three perpendicular views are
  given. The poles of the sphere are marked by arrows, and the
  equator is indicated with dashes. The six maxima along the 
  equator represent the twelve-fold symmetry;
  (c) Projection of a tcp sample obtained by cooling at
  $P\sigma^3/\epsilon=3.5$ and $k_{\rm B}T/\epsilon=0.55$. The sample has
  been expanded to $P\sigma^3/\epsilon=0.001$ and quenched to $T=0$ after
  nucleation.} 
\label{bodbcc}
\end{figure}

\begin{figure}
\centerline{\includegraphics[width=8cm]{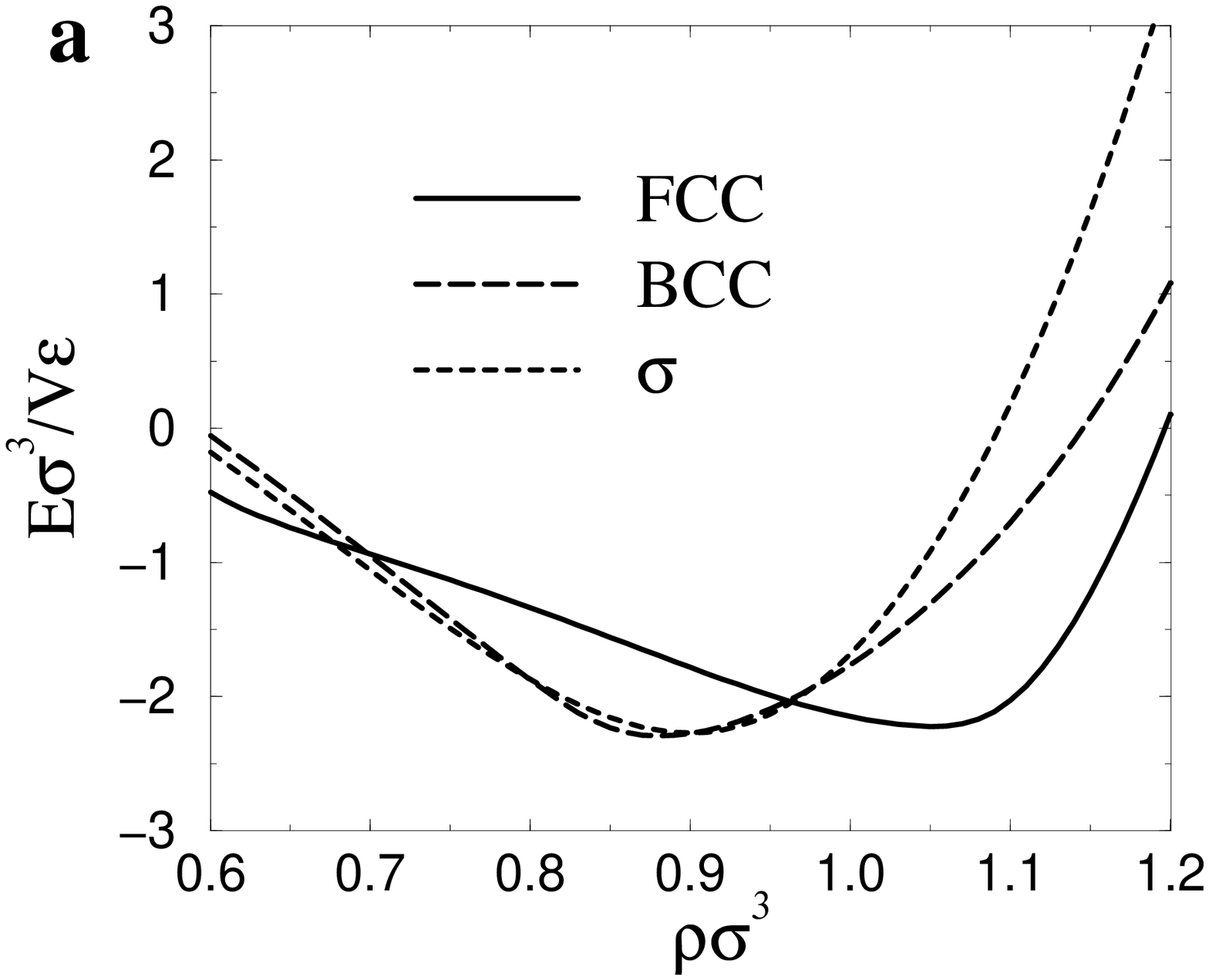}}
\vspace{1cm}
\centerline{\includegraphics[width=8cm,angle=270]{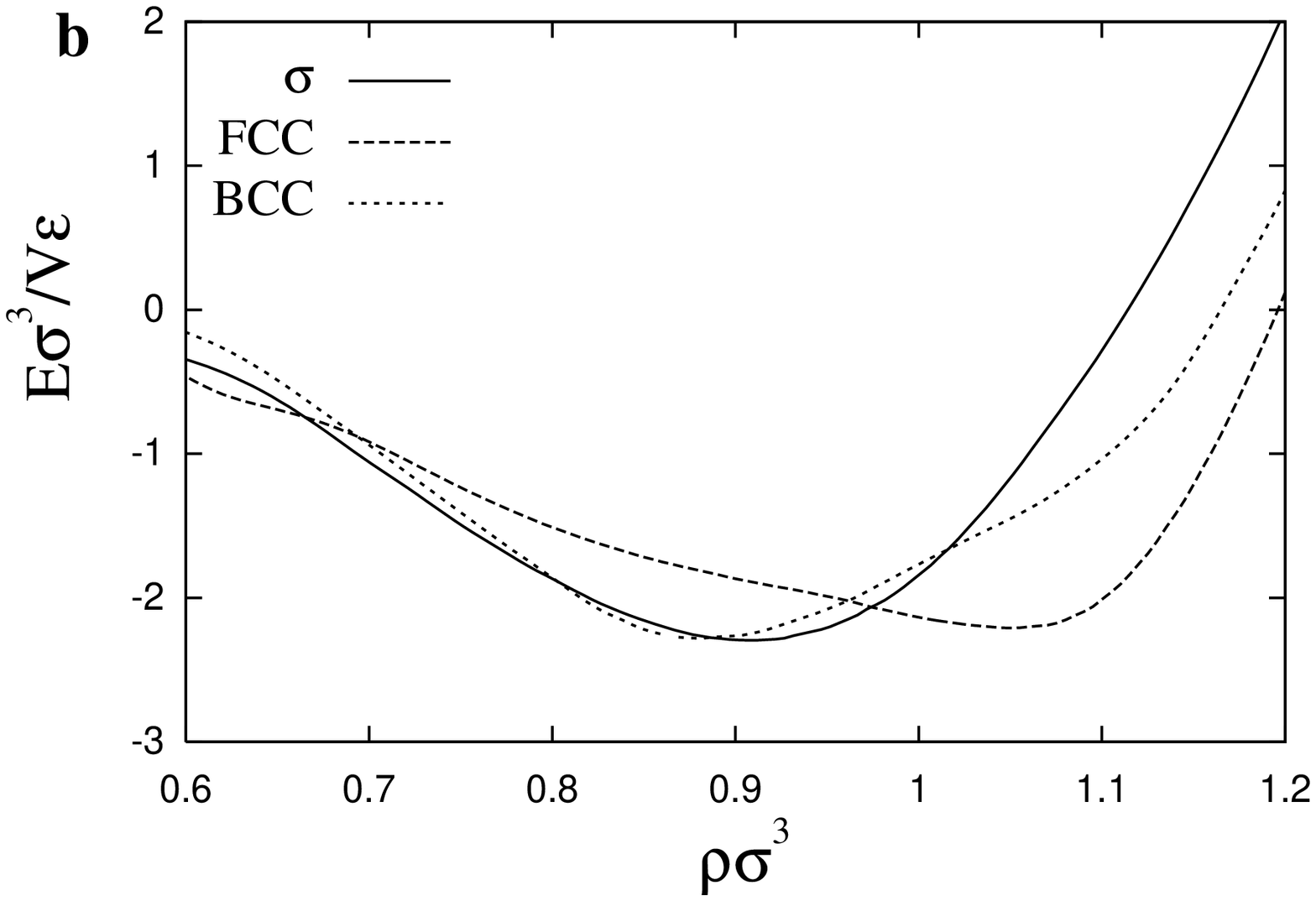}}
\vspace{1cm}
\centerline{\includegraphics[width=8cm,angle=270]{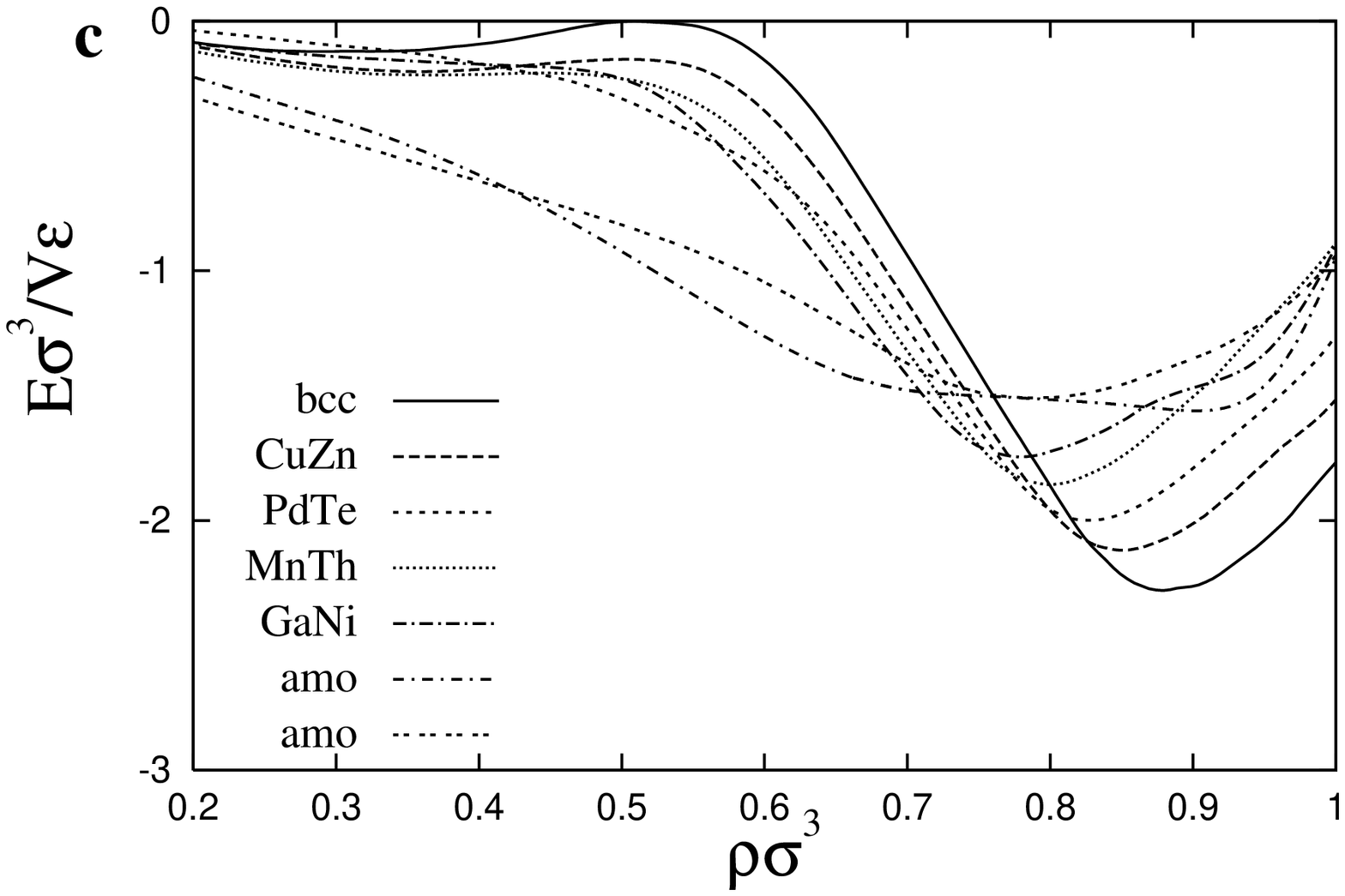}}
\vspace{0.5cm}
\caption[]{
  Ground-state energy per unit volume vs. density, in reduced units. 
  (a) Results of lattice summation of the Dzugutov pair potential for 
  ideal fcc crystal (solid curve), bcc crystal (long-dashed curve), 
  and $\sigma$-phase (short-dashed curve);
  (b) Results of MD simulation, with structural relaxation, for
  $\sigma$-phase (solid curve), fcc (long-dashed curve), and 
  bcc (short-dashed curve);
  (c) MD simulation data for the bcc vacancy phases and the low 
  density amorphous structures. The lowest minimum at the right is
  bcc followed by Cu$_{5}$Zn$_{8}$, Pd$_{4-x}$Te, Mn$_{23}$Th$_{6}$,
  and Ga$_{4}$Ni$_{3}$. The next minimum belongs to the amorphous
  phase formed from NiTi$_{2}$. The remaining double-dashed curve 
  is the amorphous phase generated by cooling the melt.}
\label{gs-energy}
\end{figure}

\begin{figure}
\centerline{\includegraphics[width=8cm]{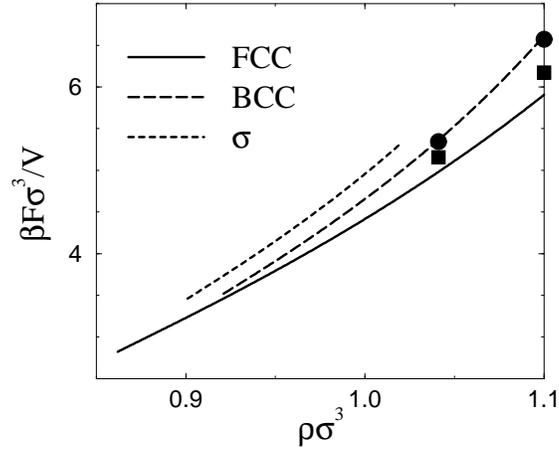}}
\vspace{0.5cm}
\caption[]{
  Free energy per volume vs. density, in reduced units, for the reference
  hard-sphere solid, computed from Eqs. (\ref{pert2}), (\ref{MWDA}), and 
  (\ref{Fid2}). Curves have the same meaning as in Fig.~\ref{gs-energy}a.
  Circular and square symbols are Monte Carlo simulation data, from 
  Ref.~\cite{CR}, for hard-sphere bcc and fcc crystals, respectively.}
\label{fvhs}
\end{figure}

For comparison with the MD simulation data, we have applied the perturbation
theory described in Sec.~\ref{theory} to predict the thermodynamic
phase behavior of the Dzugutov-potential system.  
For the fluid phase and selected solid structures, free energies were 
calculated and a coexistence analysis performed.  Our choice of structures 
was dictated by the structures actually observed in the simulations.
Figure \ref{fvhs} compares the HS part of the free energy for the fcc,
bcc, and $\sigma$-phase structures.  
Also shown, for comparison, are corresponding Monte Carlo simulation data 
from Ref.~\cite{CR}.  
From the maximum HS volume fractions of these structures -- respectively 
$0.74$, $0.68$, and $0.53$ -- stability of the HS solid is seen to be 
strongly influenced by packing efficiency.  
Therefore, at high temperatures and pressures, where entropy 
dominates the free energy and the system behaves as an effective HS system, 
the structures that are more efficiently packed are favored.  
As temperature and pressure decrease, internal energy makes an increasing
contributon to the free energy.  As illustrated by the neighbor distance 
histograms (Fig.~\ref{pot}) and the corresponding hard-sphere rdfs
(Fig.~\ref{rdftran}c), the first few coordination shells of the bcc and
$\sigma$-phase structures are more commensurate with the attractive part
of the Dzugutov potential than those of the fcc crystal, favoring these
more loosely packed structures over close-packed fcc.

\begin{figure}
  \centerline{\includegraphics[width=8cm,angle=270]{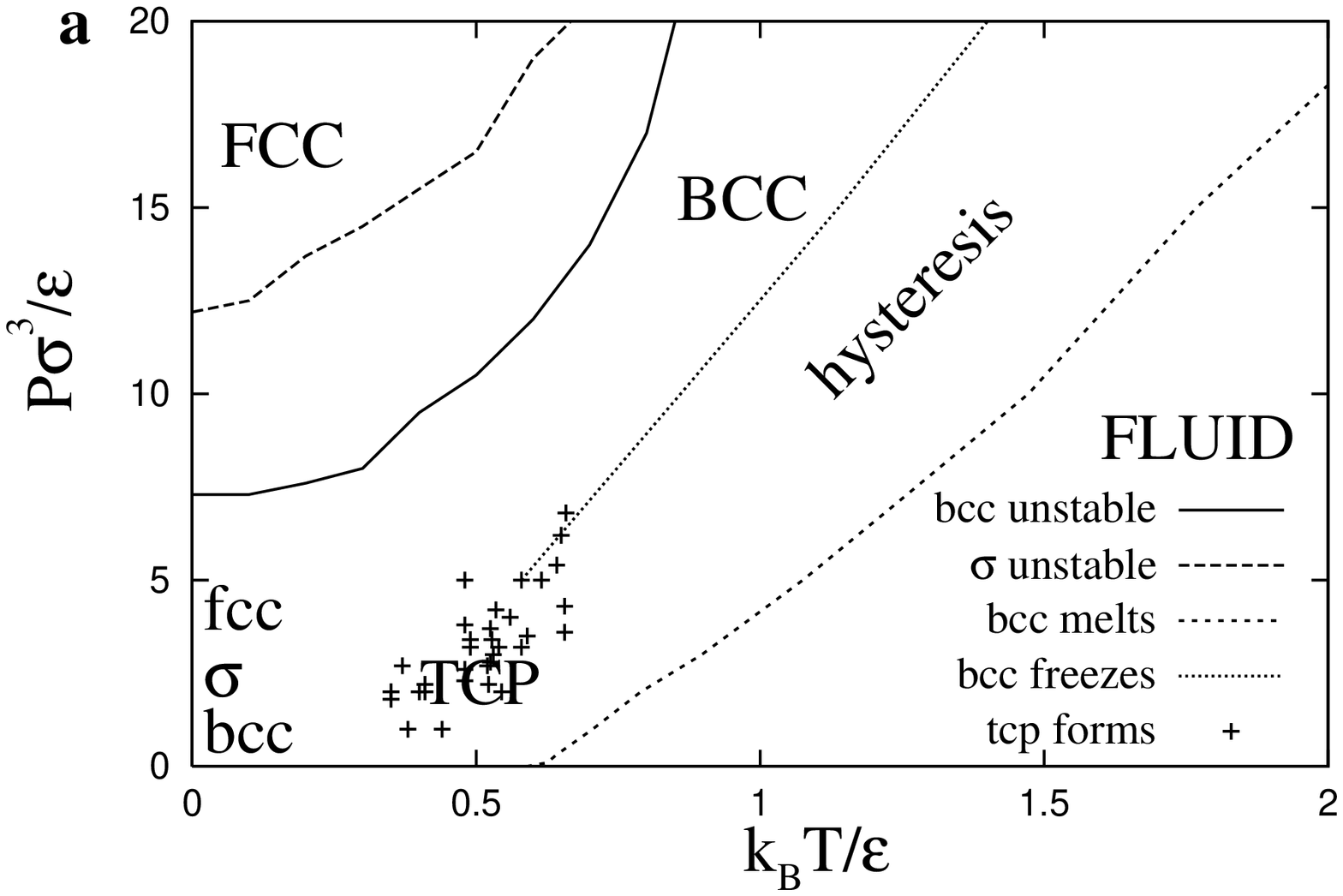}}
\vspace{1cm}
  \centerline{\includegraphics[width=8cm]{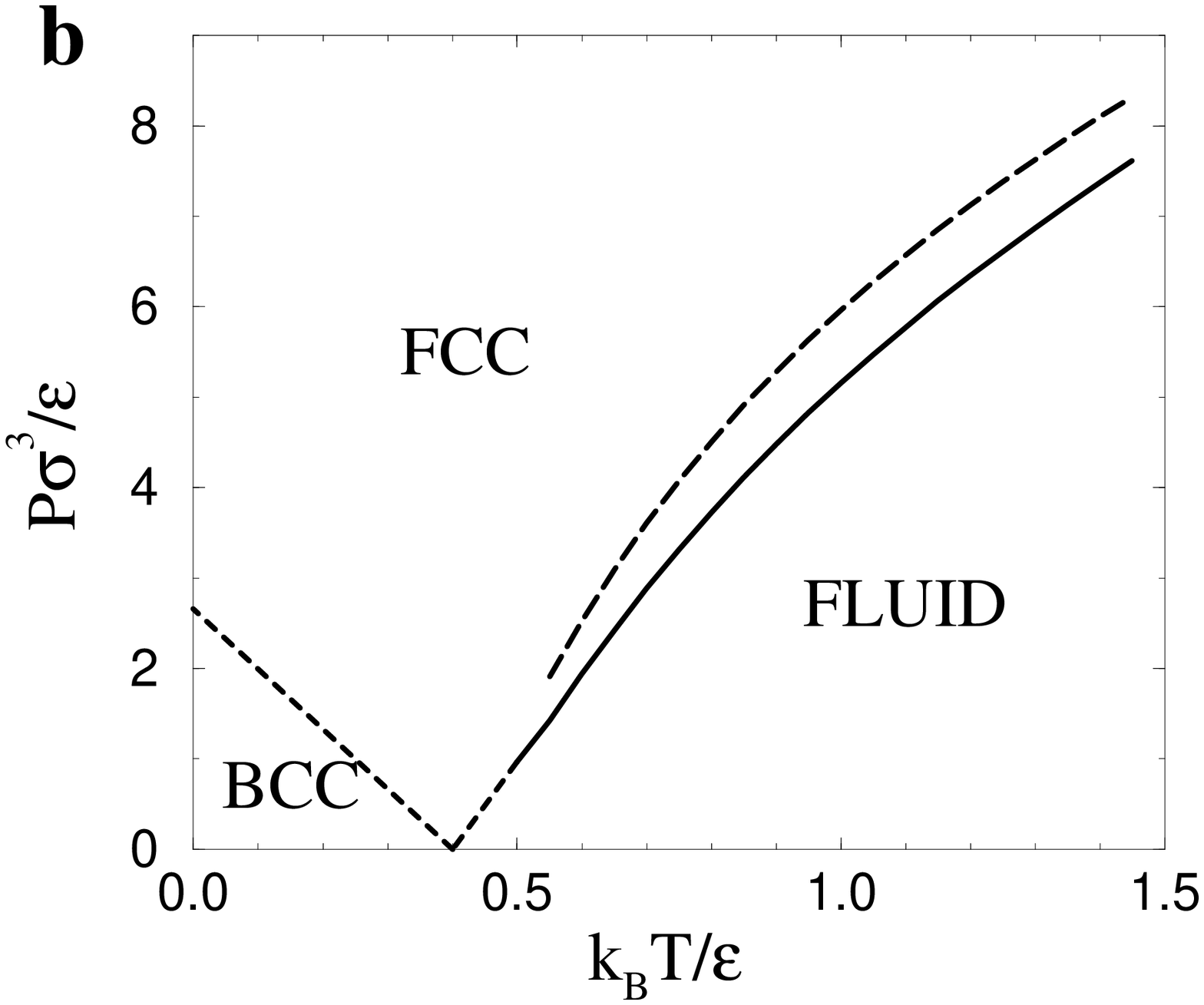}}
\vspace{0.5cm}
\caption[]{Pressure-temperature phase diagram. 
  (a) MD simulation results:  The instability lines
   denote boundaries where respective structures are destabilized 
   if compressed to high pressures. Capital letters mark phases
   formed by cooling simulations, lowercase letters phases obtained 
   by ground-state structure calculations. Crosses characterize 
   region where tcp-phase is found in cooling simulations. The region 
   between the melting/expansion transition line and the 
   cooling/compression transition line is the hysteresis region;
   (b) Perturbation theory predictions:  Phase boundaries are shown between
   fluid and fcc crystal (solid curve) and between fluid and metastable 
   bcc crystal (long-dashed curve).  Short-dashed curves are postulated 
   extrapolations to low $P$ and $T$.}
\label{phasNT}
\end{figure}

\begin{figure}
\centerline{\includegraphics[width=8cm]{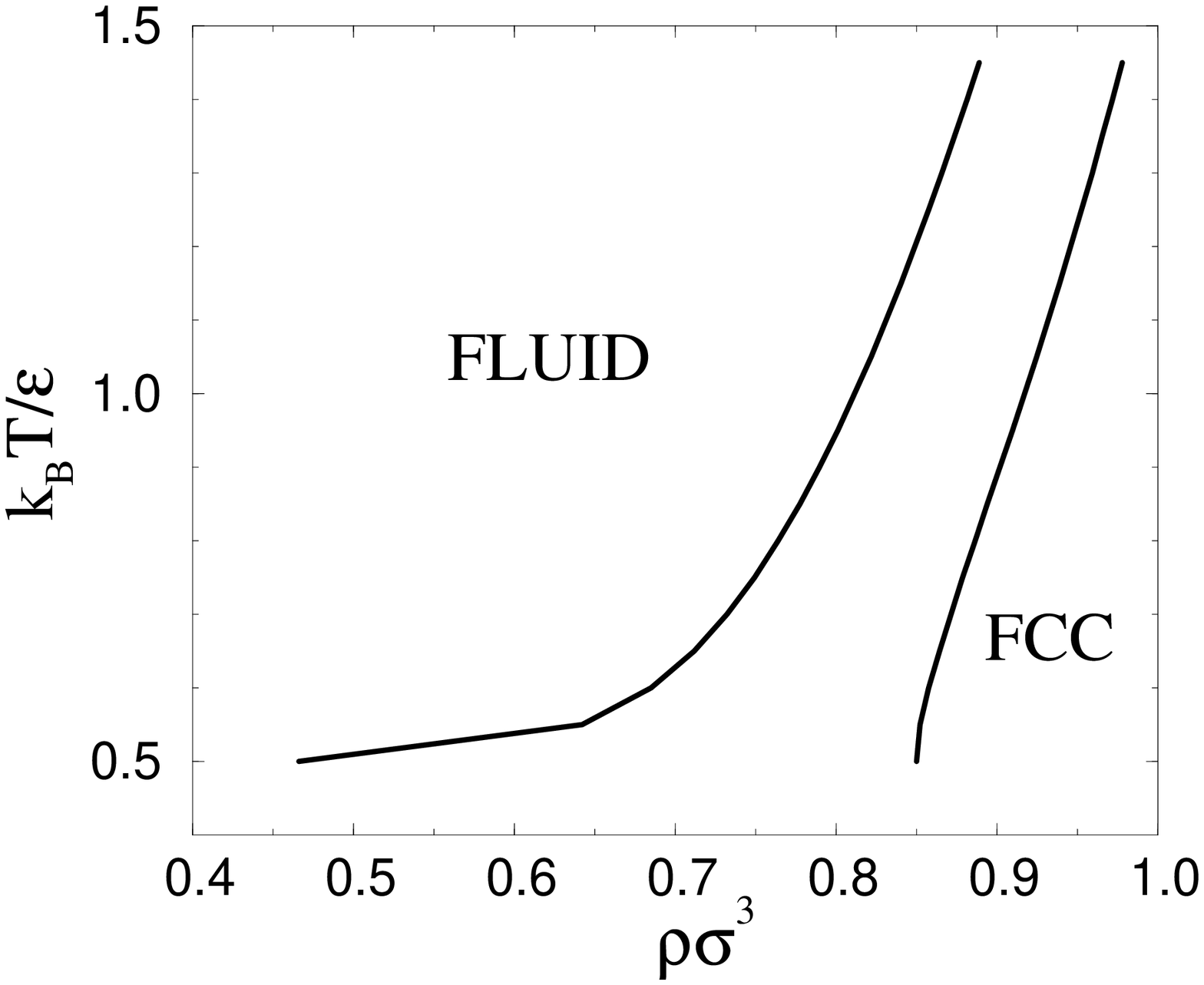}}
\vspace{0.5cm}
\caption[]{
  Predictions of perturbation theory for the fluid-solid phase diagram 
  in the temperature-density ($T-\rho$) plane.  
  For $k_{\rm B}T/\epsilon>0.5$, theory predicts the fcc crystal 
  to be the only stable solid phase.}
\label{Trho}
\end{figure}

Constructing Maxwell common tangents to curves of free energy per volume 
vs. density, thus ensuring equality of chemical potentials and pressures 
in coexisting phases, we have mapped out the phase diagram of the system.
Projections onto the $P-T$ and $T-\rho$ planes are shown in
Figs.~\ref{phasNT}b and \ref{Trho}, 
respectively.  As anticipated, the stable solid at high pressures is
the fcc crystal, while the bcc crystal is only metastable relative to fcc
(long-dashed curve in Fig.~\ref{phasNT}b).
Aside from fcc, bcc, and $\sigma$-phase, we have also considered several
tcp structures observed in the simulations and rational approximants to 
layered dodecagonal quasicrystals.
The tcp and quasicrystal structures, however, were found to be at best 
only metastable relative to the crystal structures. 
At $T=0$, lattice-sum calculations of ground-state energies
(Fig.~\ref{gs-energy}a) show 
that bcc is the stable structure for $P\sigma^3/\epsilon<2.66$.  From this
known limit, we postulate that bcc is also the stable solid structure 
at low $P$ for small but finite temperatures.  The perturbation theory 
being of uncertain accuracy for $k_{\rm B}T/\epsilon<0.5$, we further 
postulate an extrapolation of the fluid-fcc phase boundary to zero pressure.
This confines the stable bcc phase to a small pocket in the lower-left corner
of the $P-T$ diagram.

\section{Discussion and Conclusions}\label{discus}

The pressure-temperature phase diagram of the Dzugutov potential
obtained by MD simulation is surprisingly rich.
At low pressures and temperatures, the bcc-phase is stable, followed, 
with increasing pressure, by the $\sigma$-phase and by fcc 
(Fig.~\ref{phasNT}a). The bcc crystal is nucleated from the fluid 
for sufficiently slow cooling rates and sufficiently high density 
or pressure. It is also obtained by compressing the fluid. 
Below $P\sigma^3/\epsilon=5$ or $\rho\sigma^3=0.85$, tcp structures, 
including the dodecagonal quasicrystal, are formed.
The cooling scenario may be summarized as follows: 
\begin{enumerate}
\item At high cooling rates a glass is formed, which may be transformed
  into tcp or bcc solids by annealing at $k_{\rm B}T/\epsilon=0.4-0.5$.
\item At lower rates the fluid has sufficient time to reorder locally and
  crystallizes into a bcc crystal.  The bcc structure being relatively
  simple, the samples are in most cases perfect except for vacancies.
\item At sufficiently low density or pressure, a tcp structure is generated.
  Characterized by layered symmetry, the tcp-phase can be formed for 
  a wide variety of coordination polyhedra and many energetically similar
  configurations.
\item If the bcc crystal is cooled at high pressures or compressed at
  low temperatures, it transforms, at least partially, into 
  the fcc structure. 
\end{enumerate}

The stability of the lowest energy tcp-phase, namely the
$\sigma$-phase, relative to bcc, could not be determined
precisely by our simulations. Heating at low pressure shows that 
the energy and enthalpy of bcc are always lower than those of
the $\sigma$-phase. 
Comparing the two phases at higher pressures and temperatures shows that
the difference in enthalpy is no longer significant, but the energy of the
$\sigma$-phase becomes lower than that of bcc at higher pressures and
temperatures. Furthermore, upon compression bcc becomes unstable at 
lower pressure than the $\sigma$-phase.
The detailed topology of the phase diagram in Fig.~\ref{phasNT}a 
is still not completely clear and further simulations are necessary
to compute the phase boundaries exactly.  At low temperatures 
bcc appears at {\em lower} pressures than the $\sigma$-phase, 
whereas in the cooling simulations bcc is formed at
{\em higher} pressures compared to the tcp structures. It may be, 
as speculated by Dzugutov\cite{Dzugutov2}, that entropy lowers 
the free energy of the $\sigma$-phase, and especially of the quasicrystal, 
thus leading to a stable tcp or quasicrystalline state at higher temperatures.

Our theoretical calculations for a selection of perfect solid structures 
suggest that the thermodynamically stable solid phases of the 
Dzugutov-potential system are limited to fcc and bcc crystals.  
Lattice-sum calculations at $T=0$ show that the $\sigma$-phase is almost 
degenerate with, though of slightly higher energy than, the bcc crystal. 
At high temperatures ($k_{\rm B}T \gg \epsilon$), where the attractive well 
in the potential plays only a minor role, packing efficiency strongly 
disfavors the $\sigma$-phase relative to both fcc and bcc crystals.  
At intermediate temperatures ($k_{\rm B}T \simeq \epsilon$), 
perturbation theory predicts fcc and bcc crystals
to be always more stable than the relatively loosely packed $\sigma$-phase.
Thus the $\sigma$-phase appears nowhere in the $P-T$ phase diagram and
the bcc crystal appears only at low $P$ and $T$.

It may be, of course, that the first-order perturbation theory lacks 
sufficient accuracy to conclusively resolve relative stabilities of 
such closely competing phases.  In particular, fluctuations in the total 
perturbation energy, being stronger for a disorder fluid than for ordered
solids, render the theory inherently less acurate for the fluid phase.  
Moreover, the mean-field neglect of next-nearest-neighbor correlations 
in the HS rdf is less justifiable for more open structures, such as
bcc and the $\sigma$-phase, than it is for the close-packed fcc crystal.
Such correlations, if significant, would tend to lower the free energies
of the open structures and might possibly influence the order of stabilities.

Nevertheless, it should be emphasized that the predicted phase diagram 
is not necessarily at odds with the simulation data, 
if the tcp and high-$T$ bcc phases, observed in 
the simulations, are regarded as {\it metastable} with respect to fcc.
Conceivably, for kinetic reasons, the supercooled fluid first nucleates 
a metastable bcc crystallite, which upon growth transforms into 
a stable fcc crystal.  In fact, such behavior has been predicted from a 
general density-wave instability argument~\cite{Alexander}, and has been 
observed in simulations of a supercooled Lennard-Jones fluid~\cite{Frenkel}. 
Furthermore, whereas the theory has been applied to perfect structures, 
the simulations often result in solids replete with defects.  We may 
conjecture, therefore, that the defects in the $\sigma$- and tcp-phases 
observed in the simulations serve to improve the packing efficiencies 
({\it e.g.}, by increasing nearest-neighbor distances), while approximately 
preserving the average coordination distances, thereby confering energetic 
advantage over the fcc structure.

After examining a wide variety of solid structures as candidates for 
stable phases of the Dzugutov potential, we are drawn to conclude 
that only such simple structures as bcc or fcc are competitive. 
A possible exception is the $\sigma$-phase, a tetrahedrally close-packed 
structure, although one of the simplest examples of its class. 
These results appear to place the Dzugutov potential in line with 
the Yukawa, generalized Lennard-Jones, Rubidium, and Morse potentials, 
all of which favor bcc, fcc, or hcp crystals, supporting the 
general principle that simple pair potentials tend to favor 
simple structures.  
Nevertheless, it remains conceivable that quasicrystals and other 
complex structures, such as A15, Z, or H, might gain stability 
through modifications of the Dzugutov potential. 
Indeed previous work has identified somewhat related pair potentials 
for simple metals -- albeit with no counterpart in the periodic table -- 
for which stable icosahedral quasicrystals have been 
predicted~\cite{Smith,Denton-Hafner}.
Future work along these lines could examine variations of the 
Dzugutov potential in attempts to modify the relative stabilities 
of the tcp-phases and dodecagonal quasicrystals.

\section*{Acknowledgments}

We thank H.-R. Trebin for a careful reading of the manuscript and Franz
G\"ahler for supplying the data for the quasicrystal structures and
some of the square-triangle crystal structures.  ARD gratefully acknowledges
the Forschungszentrum J\"ulich for use of its computing facilities.
 
\section*{Appendix}

The following lists contain the structures investigated as possible ground 
states in the MD simulations. The notation is as found in Ref.\ \cite{Daams}.

Stacking variants of fcc(abc): hcp(ab), abcacb, abcbcb.

Frank-Kasper phases: A15 or $\beta$-W (cP8 Cr$_{3}$Si), Z (hp7
Zr$_3$Al$_4$), $\sigma$-phase or $\beta$-U (tP30 Cr$_{46}$Fe$_{54}$),
H, J, F, K', C15 (cF24 MgCu$_{2}$), T (cI162 Mg$_{32}$(ZnAl)$_{49}$), $\mu$
(hR13 Fe$_{7}$W$_{6}$), D (V$_{26}$Fe$_{44}$Si$_{30}$).

Square-triangle phases: A, H, Z, $\sigma$-phase, J, F, K', inflations of A
and Z, doubling of the tiles of $\sigma$-phase, a phase with a and h sites
mixed, tetragonal approximants of the quasicrystal with 23, 36, 172
and 836 tiles. 

Vacancy ordered phases derived from bcc: cP1, cI10 $\beta$-Hg$_{4}$Pt,
cF12 CaF$_2$, tC14 AsPd$_5$Tl, cI52 Cu$_5$Zn$_8$, cF120 Pd$_{4-x}$Te,
cF120 Sc$_{11}$Ir$_4$, cF116 Mn$_{23}$Th$_6$, cF88 Bi$_4$Cu$_4$Mn$_3$,
cI112 Ga$_4$Ni$_3$. 

The other structures are: oP12 Co$_2$Si, oP16 AlDy,
oC8 BCr, oC10 AlFe$_2$B$_2$, oC12 Ge$_2$Th, oC12 Si$_2$Zr, oC16 BCMo$_2$,
oC16 Ga$_3$Pt$_5$, oC16 HgNa, oC28 Al$_6$Mn,
oI10 B$_2$CoW$_2$, oI12 Gd$_2$Si$_3$, oI16 BMo, oI20 Al$_4$U,
oF24 Si$_2$Ti, oF48 CuMg$_2$,
tp14 Hg$_{5}$Mn$_{2}$, tP20 Al$_2$Gd$_3$, tP30 AlNb$_2$,
tI12 Al$_2$Cu, tI12 Si$_2$Th, tI16 BMo, tI28 MnU$_6$, tI32 Si$_3$W$_5$,
hP3 Cd$_2$Ce, hP3 AlB$_2$, hP5 Al$_3$Ni$_2$, hP6 InNi$_2$, hP6
CaIn$_2$, 
hR7 B$_5$Mo$_2$,
cP8 FeSi, cP20 Mn, cP39 Mg$_2$Zn$_{11}$, cP138
Al$_9$Mn$_2$Si$_{11}$, cP140 (AlSi)$_{58}$Mn$_{12}$,
cI12 Ga, cI26 Al$_{12}$W, cI58 Mn, cI76 Cu$_{15}$Si$_4$, cF96 NiTi$_2$.

\clearpage

\begin{table}
\caption{Ground-state structure and reduced density $\rho\sigma^3$ at which 
the potential energy per atom $E_{\rm pot}$ is minimal. The upper half 
contains MD results at $k_{\rm B}T/\epsilon=0.001$, the lower half 
lattice sum calculations for perfect structures.\label{enemin}}
\begin{tabular}{l|c|c}
structure & $\rho\sigma^3$ & U\\
\hline
fcc & 1.013 & -2.19 \\
bcc & 0.866 & -2.66\\
$\sigma$ & 0.879 & -2.57 \\
\hline
fcc &1.057 &-2.22 \\
bcc & 0.885 &-2.29 \\
$\sigma$ & 0.90 & -2.27 \\
\end{tabular}
\end{table}

\clearpage

\vspace{1cm}

\vspace{1cm}

\vspace{1cm}

\vspace{1cm}

\vspace{1cm}

\vspace{1cm}

\vspace{1cm}

\vspace{1cm}

\vspace{1cm}

\vspace{1cm}


\begin{thebibliography}{aa}
\bibitem{lenjon}
For the Lennard-Jones potential our simulation
methods work much better. Although it may not be possible to derive
the melting line with high accuracy, there is no problem in observing
all three phases and the transitions between them.
\end{thebibliography}
\end{document}